\documentclass[a4paper,12pt]{article}
\usepackage{graphicx,amsmath,amsfonts,amssymb,cite}
\usepackage{bm}
\newcommand{\be}{\begin{equation}}
\newcommand{\ee}{\end{equation}}
\newcommand{\ba}{\begin{eqnarray}}
\newcommand{\ea}{\end{eqnarray}}

\begin{document}

\begin{center}
{\Large\bf Parity doubling in particle physics}
\end{center}

\begin{center}
{\large S. S. Afonin\footnote{E-mail: afonin24@mail.ru}}
\end{center}

\begin{center}
{\it V. A. Fock Department of Theoretical Physics, St. Petersburg State University,\\
1 ul. Ulyanovskaya, 198504, Russia.}
\end{center}

\begin{abstract}
Parity doubling in excited hadrons is reviewed.
Parity degeneracy in hadrons was first experimentally observed
40 years ago. Recently new experimental data on light mesons caused much
excitement and renewed interest to the phenomenon,
which still remains to be enigmatic.
The present retrospective review is an attempt
to trace the history of parity doubling phenomenon,
thus providing a kind of introduction to the subject.
We begin with early approaches of 1960s
(Regge theory and dynamical symmetries)
and end up with the latest trends
(manifestations of broader degeneracies and AdS/QCD).
We show the evolution of various ideas about parity doubling.
The experimental evidence for this phenomenon is scrutinized in
the non-strange sector. Some experiments of 1960s devoted to the
search for missing non-strange bosons are re-examined and it is
argued that results of these experiments are encouraging from the
modern perspective.
\end{abstract}



\thispagestyle{empty}

\newpage

\tableofcontents
\newpage

\section{Introductory remarks}

Parity doubling in particle physics is the occurrence of
opposite-parity states of equal spin value. The problem of
parity doubling is that experimentally these states are often
approximately mass degenerate. In particle physics the phenomenon
is mainly a feature of unflavored (light non-strange) hadron spectrum.

One has always been inspired by a hope that parity doubling is
able to shed light on various knotty problems of the strong interaction
dynamics. The subject has already more than 40 years history, nevertheless
only recently two reviews appeared~\cite{jaffenew,glozrev},
which cover some related ideas in a more or less systematic manner.
We would like to make a reservation from the very outset that
the present review differs from them mainly by two aspects.
First, the issue of parity doubling in hadrons is addressed
broader, in particular, we place emphasis on the historical
development of the subject.
Second, the theoretical presentation is simplified as much as we
could with the aim of making it more readable for
experimentalists, the given review is designed in many respects
for experimentalists seeing that presently only experiment is able
to make a major contribution to clarification of modern situation
with parity doubling in particle physics.

In view of renewed interest to the parity doubling phenomenon,
a comprehensive review is certainly called for. This is, however,
a pretty formidable task which we do not pretend
to do. To a greater extend the difficulty is caused by the
fact that the ideas and approaches invoked for explanation of the
phenomenon come from quite different branches of physics, and it
is hardly possible to be a specialist in all these fields. On the
other hand, in a situation when the final truth is far from being
established in a subject, it is not easy to propose an
unprejudiced view on the subject for non-specialists. As a result,
a choice of material and references, authors' comments, etc.
can be somewhat questionable, let alone a tendency to over-concentration on
authors' personal work. In trying to escape this in our
subject, we will provide mostly a guide on the relevant literature
(in the first part of the review)
with brief explanations of proposed ideas and without giving any
preferences or criticism, an interested reader is further referred to the
original literature. In the second part of the review (Sections~6
and~7) we scrutinize experimental evidences for the parity doubling
phenomenon, discuss clustering of states near certain values of
masses and further perspectives.

The phenomenon of parity doubling in hadron spectrum has experienced
two waves of interest --- in late 1960s and in late 1990s. The
first wave was caused by the discovery of many baryon states in
1960s. The origin of the second wave (growing up to now) is more
intricate, partly it was inspired by the appearance of many
experimental data on light mesons. We will try to describe the
related ideas in a more or less chronological order.

Our discussions will concern many forgotten papers, the choice of
such a retrospective style has a motivated ground --- a known
wisdom says that the new is a well-forgotten old. We would be happy
if reading of this review stimulated someone to put forward
new ideas...

\section{1960s: Search for hidden order}

\subsection{Beginning of 1960s: The first precursors of problem}

Historically the first discovered hadron resonances gradually
formed the $J^P$
%($J$ denotes spin and $P$ does parity)
octets
$0^-$, $1^-$, $\frac12^+$, and decuplet $\frac32^+$. The minimal
group containing such representations is $SU(3)$. After
experimental establishing of these multiplets the $SU(3)$ symmetry
was finally accepted as a group of internal symmetry for
strong interactions~\cite{salam}. Nearly at the same time it
turned out that  many approaches used in that epoch for
description of strong interactions were requiring the existence of
multiplets with the opposite parity. This need was in Regge
theory (for a short review see~\cite{shirkov}), in some bootstrap
models~\cite{chan,chan2}, a bit later in the dynamical symmetry
approaches~\cite{neeman}. The proposed extensions of $SU(3)$ also
often demanded the opposite parity multiplets (see, e.g., a review~\cite{vanhove}).
The competition won the Gell-Mann's~\cite{gell}
$SU(3)\times SU(3)$ chiral symmetry\footnote{
To be precise, this is the minimal three-flavor chiral symmetry.
Say, Freund and Nambu proposed
$SU(3)\times SU(3)\times SU(3)\times SU(3)$
chiral symmetry~\cite{freund}.
The word "chiral" stems from the greek word
"$\chi\varepsilon\iota\rho$" -- "hand". In various branches of
science an object is called chiral
if it differs from its mirror image, like left and right hands.
The first systematic study of chiral symmetries in particle physics
was performed by Coleman and Glashow~\cite{coleman}.}
which gave rise to current algebra and later became an
approximate classical symmetry of Quantum Chromodynamics.
Despite the success of current algebra, at
the beginning the chiral symmetry was not widely accepted because
it predicted the opposite parity multiplets which had to be mass
degenerate with the known multiplets. This situation was far from
the experimental one, to say the least. The attitude was considerably
changed when Weinberg derived his famous formula~\cite{wein},
$m_{a_1}=\sqrt{2}\,m_{\rho}$, assuming the chiral symmetry at large four-momentum.
It became clear that the
chiral invariance can be regarded as an asymptotic symmetry of
strong interactions. This somewhat solved the problem of unwanted mass
degeneracy for parity partners.

In what follows we will often prefer to discuss the baryons and mesons
separately.

\subsection{Baryons in 1960s}

The first theoretical hints on a possible existence of parity doublets appeared
before the corresponding experimental observations. The first one was the
MacDowell symmetry~\cite{mc}: The slopes of baryon Regge
trajectories of equal isospin and signature but with
opposite parity must coincide. The second one was due to Gribov~\cite{gribov,gribo}:
The Regge trajectories of opposite parity fermions are complex conjugated.
Both results indicated that baryons must form parity doublets if the
corresponding Regge trajectories are linear.

Thus, the first explanations for parity doubling were tightly
related to the linearity of Regge trajectories motivated by the linear
dependence of hadron spin $J$ from the hadron mass squared,
\be
\label{lintr}
J=\alpha(0)+\alpha^{'} m^2,
\ee
the Chew-Frautschi conjecture~\cite{chew,chewf}. It is important to
stress that Regge theory itself did not provide convincing
arguments in favor of relation~\eqref{lintr} as this theory
establishes the fact of certain dependence of spin from the mass
squared and some restrictions on this dependence, but it does not
yield an explicit form for this dependence.
Typically the linear trajectories appear in the
relativistic descriptions while the non-linear trajectories emerge in
the non-relativistic approaches. Why do not the straight trajectories
become curved at some higher energy scale like in the
non-relativistic scattering theory based on the notion of
scattering potential? The linearity was an experimental fact, in
addition, the linear trajectories were inherent in the Veneziano
model~\cite{veneziano}, which was extremely popular at that time.
On the other hand, this model had problems with the incorporation
of the MacDowell symmetry. The universal slope of Regge trajectories
$\alpha^{'}$ (of the order of $1$~GeV$^{-2}$) is naturally
related to the universal range of strong interactions (of the order of
$10^{-13}$~cm). If the trajectories are curved at much higher
energy scale, this means then that strong interactions have an
additional characteristic scale. In this case one observes the
linear trajectories simply because every curve looks as a
straight line at sufficiently small interval. If it were the case,
the self-consistency of the analytical $S$-matrix approach would be
questionable (say, one of postulates of the $S$-matrix theory is
decomposability of the $S$-matrix due to finite range of strong
interactions). There were proposals that this scale at high
energies (high compared to the known resonance region) could be provided
by quark masses as long as quarks were very heavy in the old quark
models, of the order of $5$~GeV or more. Later a more convincing
justification for the linearity of trajectories was proposed ---
the relativistic hadron strings. But this is out of the scope of
our topic.

In several years parity doubling among some nucleon resonances was indeed
observed experimentally. An "explosion" of these observations happened in
1967 (see, e.g.,~\cite{don,don2,don3,don4,don5}). Barger and Cline~\cite{bar,barg} immediately attributed
the phenomenon to a manifestation of the MacDowell symmetry.
%(in~\cite{iw} it was argued, however, that there is no relation between
%the MacDowell symmetry and the existence of parity doublets).
However, along with the parity doublets one observed some notable
parity singlets, e.g., the ground states. This fact seemed to
contradict the MacDowell symmetry and caused much discussions.
Different ways out were proposed, for instance, vanishing residues
for the corresponding parity partners~\cite{james}, but such {\it ad hoc}
solutions did not seem to be satisfactory~\cite{shirkov}.
Different authors tried to adjust the situation in the framework of representations
of the Lorentz group or its extensions (see, e.g.,~\cite{barklein,kleinert,tol,toll,mukunda,dkm}).
The proposed schemes indeed required the parity duplication of some
baryons since they were (partly) based on the Toller analysis~\cite{tol,toll}. In
Toller's scheme one assigns hadrons (in the rest frame) to
irreducible representations $(j_1,j_2)$ of the Lorentz group, then one
considers the "Toller" quantum number
\be
\label{toln}
M=|j_1-j_2|.
\ee
The states with $M=0$ are parity singlets while the states having
$M\neq0$ are parity doublets. Inasmuch as baryon spin $J$ is
half-integer and $|j_1-j_2|\leq J\leq j_1+j_2$, the pair of
indices $(j_1,j_2)$ has to consists of one integer and one
half-integer numbers, hence, all baryons transforming under the
representation $(j_1,j_2)$ are parity doubled.

At the same time it was realized that parity doubling in the Regge
theory is a particular solution for the so-called "conspiracy"
among different Regge trajectories (see, e.g.,~\cite{domokos} for
references): In order to avoid kinematic singularities of
invariant amplitudes at vanishing momentum transfer, some linear
combinations of certain partial wave amplitudes have to be equal
to zero~\cite{gribov2}. This problem emerges when one takes into
account the spin of particles and differences in masses.
Generally speaking, a solution of the conspiracy
problem is not unique. Consequently, a natural question emerged,
why parity doubling is preferred? Various proposals appeared
that this is a consequence of $SO(4)$ space-time symmetry of
scattering amplitude at vanishing momentum transfer
(see, e.g.,~\cite{tol,toll,sawyer,macfadyen,tol2} and references therein),
for some dynamical reasons one also observes an imprint of
this symmetry at non-vanishing momentum transfer. The
Lorentz invariance (or $SO(4)$ after the Wick rotation) of
scattering amplitude was argued to result in the
existence of "daughter trajectories" for any Regge trajectory
(earlier this result was deduced from the analyticity properties of
scattering amplitude) and
in the appearance of parity doubled type of conspiracies. Extending
$SO(4)$ by parity, one thus can conclude that parity
doubling is a consequence of $O(4)$ symmetry of the spectrum.
However, a certain care must be exercised thereupon. The
invariance group of a scattering amplitude need not coincide with
the classification group for its spectrum of the bound states.
The coincidence takes place for a scattering
amplitude with all the external particle masses equal~\cite{domokos}.

We shortly remind the origin of ideas related to the $O(4)$
symmetry for Regge theory. In 1954 Wick~\cite{wick} introduced
his famous "rotation" from Minkowski space to Euclidean one. It
was proposed for mathematical simplification of the Bethe-Salpeter
equation. Cutkosky~\cite{cutkosky} immediately made use of Wick's
trick to find a complete
set of bound state solutions in the case  of the Bethe-Salpeter
equation for two scalar particles. The degeneracy of solutions
turned out to be identical to that of the nonrelativistic hydrogen
atom. The method itself happened to be, in a sense, dual to
Fock's treatment of hydrogen atom~\cite{fock} where the $O(4)$
symmetry is manifest. In ten years Domokos and Sur\'{a}nyi~\cite{domokos2}
noted that such a higher symmetry implies interesting consequences
for Regge trajectories. They found that every singularity in the
angular momentum plane induces a series of other singularities of
the same nature following the original one at unit steps. This
situation is a natural consequence of $O(4)$ symmetry: There is,
in fact, one singularity in the complex $O(4)$ angular momentum
variable, which generates the series of singularities above when
one decomposes according to the usual $O(3)$ angular momentum.
Stated differently, one four-dimensional pole is equivalent to a
superposition of poles in the usual three-dimensional angular
momentum plane. In that way the daughter trajectories emerge. The
$O(4)$ theory of Regge trajectories was further elaborated by
Freedman and Wang~\cite{freedman,freedman2,freedman3,freedman4}. In particular, they examined
the reason of Coulomb degeneracy in Bethe-Salpeter models. The group
$O(3)$ of three-dimensional rotations is the invariance group of
Bethe-Salpeter equations for nonzero total energy as the total
energy-momentum four-vector is fixed under $O(3)$ rotations.
For zero total energy, however, this four-vector vanishes, and the
equation becomes invariant under $O(4)$ transformations of its
integration variables. This very extra degree of invariance
ensures the existence of daughter trajectories in much the same
way that the extra degree of invariance for some infinite range
potentials ensures the Coulomb degeneracy of bound states. As a
byproduct, the higher symmetry (with the ensuing decomposition
of amplitudes
in $O(4)$ harmonics) automatically resolved a long-standing
problem with the ambiguity of the asymptotic behavior of the
unequal-mass scattering. Although in general case (unequal mass scattering)
$O(4)$ is not an
exact symmetry of the scattering amplitude, this higher symmetry
can be a good symmetry for the spectrum of the amplitude, at least
in the first approximation. For this reason the spectrum of $\pi N$ resonances should
follow the underlying higher symmetry. This point was scrutinized by
Domokos~\cite{domokos3}.

Let us present the key features of $O(4)$ partial-wave analysis.
One decomposes an amplitude in the four-dimensional spherical
harmonics,
\be
\label{4h}
{Z_n}_l^m(\beta,\theta,\phi)=p_{nl}(\beta)Y_l^m(\theta,\phi),
\ee
where $Y_l^m(\theta,\phi)$ is a usual three-dimensional spherical
harmonic, $n=0,1,2,\dots$ is analogous to the principal quantum
number in the hydrogen atom, and $p_{nl}(\beta)$ can be
expressed through Legendre or Gegenbauer functions of $\cos\beta$,
which gives the restriction $l\leq n$. The spectrum (both poles and
branch cuts) appear as simple singularities in the $n$ plane, in
the $l$ plane it shows the pattern required by the higher symmetry
$O(4)$. Then one introduces the integer quantum number $\kappa$,
\be
\label{4h2}
n=l+\kappa.
\ee
It is called "relative-time parity" and bears a close analogy with
the radial quantum number in the hydrogen atom. The even values
of $\kappa$ give rise to the daughter Regge trajectories. The odd
values do not correspond to observable particles (the odd-$\kappa$ poles
in the physical region violate unitarity). Thus, starting, e.g.,
from a parent trajectory with the states at $l=0,2,4,\dots$, one
obtains the daughter states corresponding to the even valued
$O(4)$ spherical harmonics.

The $O(4)$ partial-wave analysis may be regarded as a particular
realization of generalized partial-wave analysis concept for the
$S$-matrix, which was put forward by Salam and
Strathdee~\cite{salstr}. According to this concept, the
partial-wave analysis can be probed by almost any complete set of
orthogonal functions and if a certain choice turns out to be
successful phenomenologically and the corresponding set realizes a
representation of some higher symmetry group, then the
corresponding higher symmetry is a good candidate for the
underlying dynamical symmetry generating the observed spectral
recurrences. The concept was illustrated in~\cite{salstr} by
decomposition in the $O(6)$ spherical harmonics. Being isomorphic
to Wigner's higher symmetry $SU(4)$, the group $O(6)$ was assumed
to include internal symmetries.

The previous note makes a bridge to another approach to the description
of parity doubling --- the dynamical symmetry formalism.
By a dynamical symmetry group one means here a group which gives
the actual quantum numbers and degeneracy of a quantum-mechanical
system (sometimes it is called "hidden", "accidental" or
"spectrum-generating" symmetry). In this approach symmetries of
Hamiltonian do not play an important role. Physically the
dynamical group reflects the internal structure of the system.

Let us explain the idea by a classical example --- the hydrogen atom (H).
It has the $O(3)$ rotational invariance, hence, each state of discrete
spectrum can be labelled by $|lm\rangle$, where $l$ and $m$ are
the usual angular momentum quantum numbers --- the angular momentum and
its projection. However, as was first
discovered by Fock~\cite{fock}, the actual symmetry of discrete
spectrum for the H-atom
is $O(4)$. It is manifested by the existence of the principal
quantum number $n$ numerating the energy levels,
\be
\label{hydsp}
E_n\sim\frac{1}{n^2},\qquad n=l+n_r+1,
\ee
where $n_r$ is the radial quantum number. As a
consequence, the discrete states of H-atom are labelled by three
numbers, $|nlm\rangle$. All wave functions corresponding to states
with the same energy, i.e. labelled by the same $n$, fall into one
irreducible representation\footnote{Although $O(4)$ has two Casimir
operators, i.e. irreducible representations are labelled by two
indices $(j_1,j_2)$, one index is enough for labelling of
irreducible representations in the Coulomb (Kepler) problem.
The reason is that the Casimir operators happen to be equal in the
case of the Coulomb potential, hence, only the representations with
$j_1=j_2$ are realized in nature.}
of $O(4)$~\cite{bargmann}.
In thirty years Malkin and Man'ko made the
next breakthrough in the group theory of H-atom~\cite{malkin,malki}: the
full dynamical symmetry group is the conformal group $O(4,2)$
which includes $O(4)$ as a maximal compact subgroup.
Soon alternative derivations
of this result were proposed (see references in~\cite{malkin2,barklein2}).
It turned out (see references in~\cite{barklein3}) that all states
of discrete spectrum as well as the continuum spectrum and all
radiative transitions can be compactly described within the
$O(4,2)$ dynamical group, i.e. the whole relativistic theory of
H-atom (without account of electron spin) can be formulated in
terms of this group, with the $O(4,1)$ subgroup being the dynamical
group of the bound states and the $O(3,2)$ that of the scattering
states. This is tightly related with the fact that the Kepler
problem can be formulated as $O(4,2)$ dynamical group
theory~\cite{gyoergyi}.

What is the physical meaning of $O(4)$ and $O(4,2)$ dynamical symmetries for
the H-atom? The $O(4)$ symmetry tells us that if we know a wave
function of state with a given energy then acting by generators of
$O(4)$ on this wave function we are able to obtain the wave
functions of all states with the same energy without solving the
Schr\"{o}dinger equation. The larger $O(4,2)$ symmetry\footnote{More exactly,
its $O(4,1)$ subgroup when discussing the discrete spectrum of H-atom.
The dynamical group $O(4,1)$ connects states with different principal
quantum numbers $n$ and contains $O(4)$ as a subgroup.
The totality of all the bound-state wave functions
carry a representation of $O(4,1)$.} tells us that by applying the
same procedure we will get the whole set of wave functions for
discrete spectrum.

The success of dynamical symmetry approach in the H-atom
%(and in the Kepler problem~\cite{barut})
inspired
to apply similar ideas to hadron physics. It was assumed that the
quantum theory of hadrons can be formulated in terms of
irreducible representations of some dynamical groups (both compact and
noncompact) with no Hamiltonian or space-time
coordinates at all. The problem was to identify an appropriate
dynamical group and find its relevant irreducible representations.
Indeed, in the usual dynamical approach one finds a discrete spectrum by
solving an eigenvalue equation. On the other hand,
if one knows all solutions of an eigenvalue equation one can
always assign the corresponding eigenfunctions to one irreducible
representation of some group (at least for the differential eigenvalue equations).
In this sense a search for the solutions of dynamical equations
might be equivalent to a group-theoretical search for higher symmetry.

The experimental spectrum of baryons happened to be qualitatively similar to
that of the H-atom. This observation inspired Barut
{\it et al.}~\cite{kleinert,barklein3,barut,bck,bck2} to apply the dynamical
$O(4,2)$ group to description of baryons. The unitary irreducible
representations of $O(4,2)$ contain the states which for given
quantum numbers are characterized in the rest frame by
$|njm,\pm\rangle$. Here $\pm$ refers to the parity determined from
the parity of the ground state. There are two possible ways of
parity doubling in the $O(4,2)$ representations. In the first case
all states have their opposite parity counter-part. In the second
case all states for a given $n$ are parity doublets, except one
parity singlet state emerging at $j=n-1$ (see Fig.~1).
The latter case is realized in the H-atom, it seemed to be
preferable also for nucleons.
The obtained accordance with the experimental data
(both on mass spectrum and on formfactors) was rather
encouraging.

\begin{center}
\begin{figure*}
\vspace{-4cm}
%\hspace{-1cm}
\includegraphics[scale=0.8]{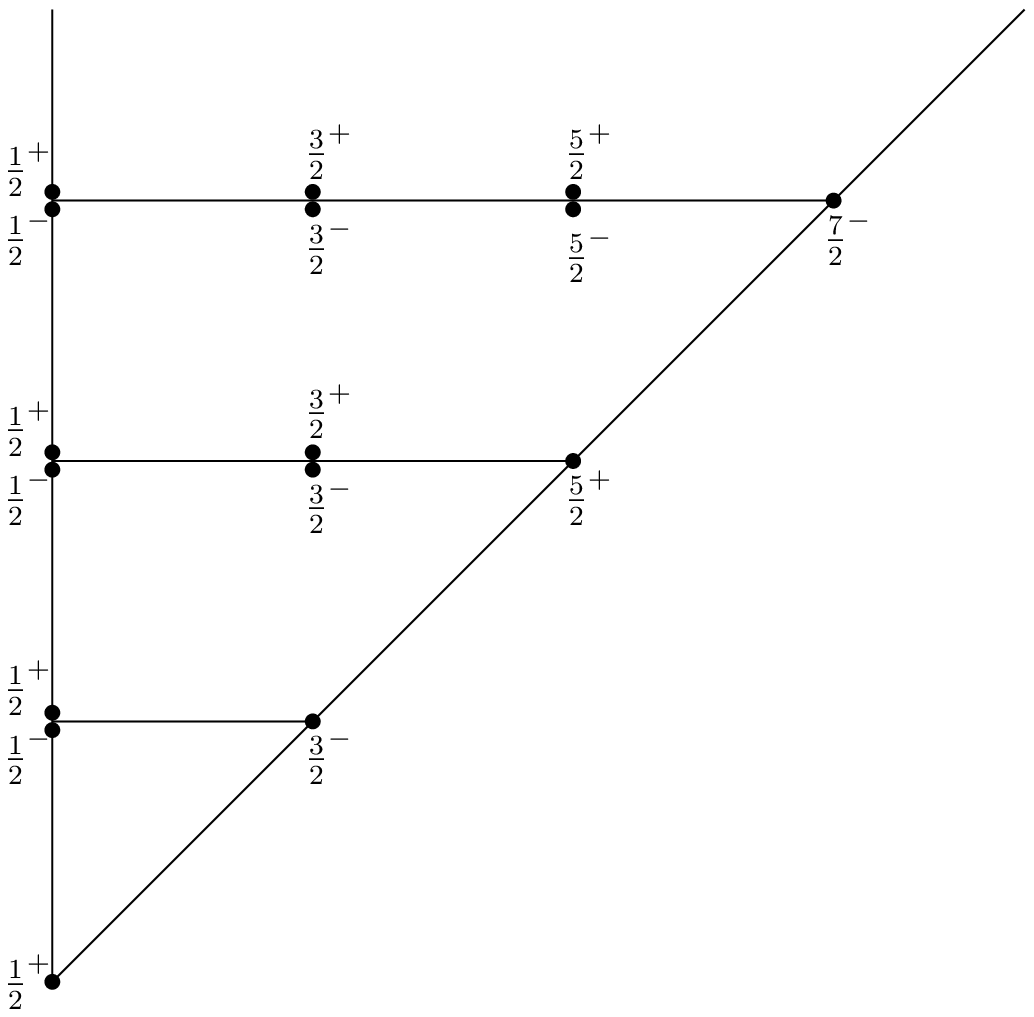}
\vspace{-12cm}
\caption{\small The weight diagram of the hydrogen-like
$O(4,2)$-representation for the nucleon $J^P$ states
(a simplified figure from~\cite{barut}).}
\end{figure*}
\end{center}

\vspace{-1.3cm}

Originally the dynamical symmetry approach was introduced to hadron physics
independently of the group theory for the H-atom. The
corresponding ideology was formulated by Dashen and
Gell-Mann~\cite{dashgell}. A general scheme for accommodation of
states with different parities was discussed in~\cite{dothan}. In
short, one deals with a finite number of energy levels (hadron masses) in hadron
physics. Before those papers, the situation was usually accommodated by a
finite-dimensional irreducible representations of compact groups,
like $O(4)$ in the H-atom. However, if there are many states,
an infinite sequence of discrete energy levels can be a
permissible idealization. In this case the use of an
infinite-dimensional representation can turn out to be a more
effective approximation than the use of a finite-dimensional one.
A group possessing such a unitary irreducible representation has
to be noncompact. After this justification, the use of noncompact
dynamical groups became quite popular, the conformal group
$O(4,2)$ is an example.

The program for determining the whole hadron mass spectrum and
formfactors with the help of some underlying dynamical group was very
ambitious, the peak of activity occurred in 1967-1968. Finally the program
failed, the number of papers on the spectrum-generating approach
decreased exponentially, although this method was not forgotten
completely (see, e.g., a classification of meson Regge
trajectories based on the $SO(4)$ dynamic symmetry in~\cite{imz}).

At that time the success of current algebra and partially conserved
axial-vector current hypothesis made apparent the
fact that strong interactions are approximately symmetrical under the
$SU(3)_L\times SU(3)_R$ chiral group~\cite{gell} and, hence, all hadrons
should fall into multiplets of chiral group (see, e.g., the related
discussions in~\cite{gh,wein2}) containing
degenerate states of positive and negative parity.
This symmetry (more precisely, its Wigner-Weyl realization) is broken to
the vector $SU(3)_V$ subgroup and the broken part of the chiral symmetry manifests
itself through the appearance of eight nearly massless Goldstone bosons.
In other words, the chiral symmetry is realized in the Nambu-Goldstone mode.
In 1969 Dashen noticed~\cite{dash}, however, that the residual symmetry of hadron
spectrum could be $SU(3)_V\times \mathcal{Z}$, where $\mathcal{Z}$ is a
discrete symmetry, which leaves the vacuum invariant and leads to parity
doublets. Namely, the discrete group $\mathcal{Z}$ consists of six elements
$\{1,P,Z,Z^{\dagger},PZ,PZ^{\dagger}\}$, where $P$ is the parity operator
and the discrete operation $Z$ is related to the axial hypercharge $Y_5$:
$Z\equiv e^{i2\pi Y_5}$. The group $\mathcal{Z}$ has two one-dimensional
representations that are parity singlets and one two-dimensional representation
which contains the states with opposite parities, the latter representation exists
only if $n_f>2$. The particles will then fall
into multiplets corresponding to one of these irreducible representations.
In the second case one must observe parity doubling in the mass spectrum.
Within this picture all states on a given Regge trajectory must be
either parity singlets or doublets.
The related phenomenology was occasionally appearing in the
literature.
In thirty years, however, this possibility was excluded by the rigorous QCD
inequalities~\cite{kks}.

A few years later, in 1973, the fundamental theory of strong interactions,
QCD, was introduced~\cite{fritzsch} after the discovery of its asymptotic freedom~\cite{asfr,asfr2,asfr3,asfr4}
and many theoreticians switched over QCD. Nevertheless, QCD was not shedding light on
the problem of parity doubling for a long time. Meanwhile,
experimentalists were discovering and confirming more and more new parity doubles
in the baryon sector...

\subsection{Mesons in 1960s}

Because of a shortage of experimental data the story in the meson sector is not so rich.
The same as in the baryons the first arguments were based on Regge theory
and on the dynamical group approach.

Barger and Cline~\cite{bar2} associated the absence of
backward peaks in $\pi^+\pi^-$, $\pi^+K^-$, $K^+K^-$, and
$\bar{N}N$ elastic scattering with the occurrence of meson
resonances in highly correlated sequences of angular momentum
states with alternating parities called "towers". The first $J^P$
tower is $(0^+,1^-)$ (of both isospin), the second one is
$(0^+,1^-,2^+)$, the third one is $(0^+,1^-,2^+,3^-)$, etc.
(see Fig.~2). According to modern knowledge, Regge trajectories of
different isospin, the $(\omega,\rho)$ and $(f_2,a_2)$ trajectories
in our case, are practically degenerate due to a negligible admixture of
strange quark. Experimentally the four trajectories
$(\omega,\rho,f_2,a_2)$ coalesce into one master trajectory, in
Regge theory this fact is known as exchange
degeneracy\footnote{Exchange degeneracy is the approximate dynamical
degeneracy of two sets of trajectories with opposite signature and
$G$ parity, e.g., the $\rho$ and $a_2$ trajectories.
Using the Mandelstam variables $(s,t,u)$,
exchange degeneracy originates from the absence of contribution of $u$
channel resonances to an amplitude $A(s,t)$. Like linearity of
trajectories, exchange degeneracy does not rigorously follow from
Regge theory, it was a feature of the Veneziano
model~\cite{veneziano} and was explained by the old quark
hypothesis: The exchange forces stem from direct interaction
between heavy quark and antiquark, the exchange mesons cannot be
lighter than quarks, hence, exchange forces are very short-range,
i.e. negligible. Exchange degeneracy was first proposed by Arnold
by analogy with potential theory~\cite{arnold}.}.
The tower hypothesis predicted for linear rising meson trajectories
the existence of large number of meson states in the mass regions
called $R(\sim1700\,\text{MeV})$, $S(\sim1930\,\text{MeV})$,
$T(\sim2100\,\text{MeV})$, and $U(\sim2300\,\text{MeV})$. In
addition, in order to build up full nucleon-antinucleon elastic
scattering amplitude one required a strong local parity degeneracy
of the meson states of the kind that towers could provide. Making
use of the fact that $\bar{N}N$ inelastic cross section is bigger
than the elastic one, it was concluded that mesons should be
strongly coupled to the $\bar{N}N$ annihilation inelastic
channels, hence, the discovery of many meson resonances in
$\bar{N}N$ annihilation would provide a crucial test for the tower
hypothesis. In about thirty years all these conclusions were
qualitatively confirmed by the Crystal Barrel experiment on $\bar{p}p$
annihilation in flight~\cite{anietall,anietall2,anietall3,anietall4,ani,ani2,bugg}. It is quite
remarkable that a recently obtained preliminary picture of non-strange
meson spectrum (see Fig.~8) had been qualitatively anticipated in
the preQCD time.

\begin{center}
\begin{figure*}
\vspace{-4cm}
%\hspace{-1cm}
\includegraphics[scale=0.8]{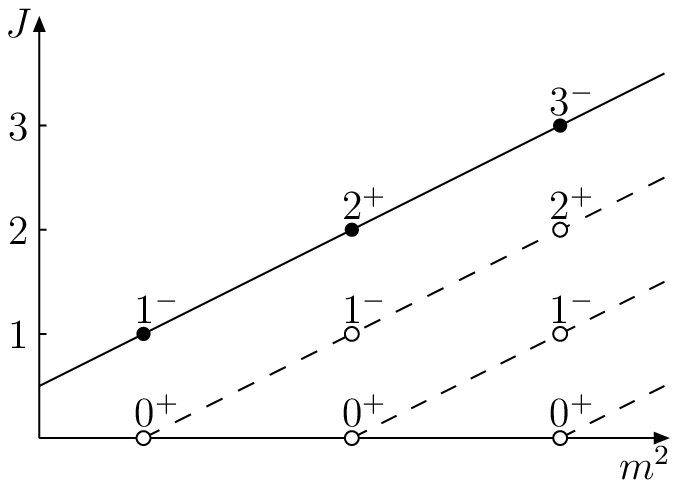}
\vspace{-17cm}
\caption{\small Linearly rising master trajectory (solid line and
filled circles) and the associated towers of meson states. The
states on the daughter trajectories (dashed lines) are denoted by
open circles.}
\end{figure*}
\end{center}

\vspace{-1.06cm}

A bit earlier Barut~\cite{barut3,barut3b} applied to the non-strange mesons
the hydrogen-like description based on the $O(4,2)$ dynamical group.
As a result, a similar picture of meson "towers" emerged. Say, the
states in the pion towers of $O(4,2)$ are
$(0^-)^{n=1}$; $(0^-,1^+,2^-)^{n=2}$;
$(0^-,1^+,2^-,3^+)^{n=3}$; $\dots$\,. The states belonging to
the same tower are naturally degenerate because they have equal
"principal" quantum number $n$. The parity conjugated towers "grow
out" of the $\rho$-meson. The same as in the baryon case, there
are two ways of parity doubling within the rest-frame dynamical
group $O(4,2)$ --- either with parity doubled ground state or with
parity singlet one.

An interesting
proposal made Alessandrini~\cite{alessandrini}. He tried to apply
to mesons the Gribov's mechanism of parity doubling for the
fermion Regge trajectories~\cite{gribov,gribo} (the parity-doubled conspiracy
at zero momentum transfer). It was shown that this can be achieved
if the wave functions of spin $J$ bosons belong not to the $\left(\frac12J,\frac12J\right)$
irreducible representation of the Lorentz group (because $M=0$ in
this case, see Eq.~\eqref{toln}) but to the
reducible representation
\be
\label{alex}
\left[(1,0)\oplus(0,1)\right]\times\left[\frac{J-1}{2},\frac{J-1}{2}\right].
\ee
This representation is analogous to the Rarita-Schwinger
representation for half-integer spin.

Domokos {\it et al.}~\cite{dkm} introduced a certain complex extension
of Lorentz group (isomorphic to the "chiral" Lorentz group
$SL(2,C)\times SL(2,C)$). Covariance under this group requires
parity doubling not only for baryons, but also for mesons with
$M\neq0$ if the corresponding Regge trajectories are linear.

The exercises above looked rather academical as experimentally
there was no example of degenerate parity partners in the mesonic sector.
In addition, making use of the fact that residues of states with $M=0$ and
$M\neq0$ behave differently at vanishing momentum transfer, the analysis of
various helicity amplitudes showed that all known mesons had $M=0$, i.e.
they are parity singlets if the Toller theory is correct\footnote{In fact,
at that time the well established mesons were either the Goldstone bosons
or belonged
mostly to the principal Regge trajectories. According to the modern
experimental data, all meson states on these trajectories are indeed
parity singlets, this will be concerned in Sections~6 and~7 (see Fig.~8).}.
Just as in baryons, after advent of QCD the problem of parity
doubling was forgotten for a long time.

\section{End of 1980s: Revival of interest}

In the late 1980s the problem was risen again by
Iachello~\cite{ia}. The purpose of his work was
"... {\it to bring attention to a major feature of baryon spectra that,
although extensively investigated in the late 1960s within the context
of both chiral symmetry~\cite{pagels} and Regge-pole theory~\cite{collins},
has, in recent years, been somewhat overlooked. This is the occurrence
in the spectra of parity doublets}..." It was argued that the
occurrence of parity
doubling in some states and nonoccurrence in others are a
consequence of the geometric structure of hadrons, i.e. the
underlying dynamics is similar to that of some molecules.
To reveal the underlying physics Iachello proposed
some baglike and stringlike models. We will describe the idea
confining ourselves to the string case only.

Consider the following model for baryon: Three identical quarks
are on the tops of Y-type string (the Mercedes-Benz type of string
junction, the picture resembles the ammonia molecules $NH_3$ where
parity doubling of energy levels is known to occur). Aside from
the rotational invariance, there is a symmetry with respect to the
permutations of quarks. This symmetry is isomorphic to the
geometric point group $C_{3v}$. In the theory of point groups, the
symbol $C_n$ denotes the symmetry with respect to rotation on the
minimal angle $2\pi/n$. This angle is equal to $120^{\circ}$ in our
case, hence, $n=3$. The symbol $v$ means the reflectional symmetry
with respect to the vertical plane. Consider the mesons. They are
made of quark and antiquark attached by a string. Since quark and antiquark
are not identical particles, this system has the geometric
symmetry $C_{\infty v}$ (the same as the symmetry of linear
molecule). The transformations of this group consist in rotations
and reflections on a plane,
i.e. it is isomorphic to $O(2)$. At enlarging angular momentum $l$
the Y-string produces an elongated shape, i.e. it becomes
reminiscent of the quark-diquark structure. Thus, at large $l$ the
geometric group of baryons $C_{3v}$ converts into $C_{\infty v}$.
The discrete group $C_{3v}$ has two one-dimensional
representations, called $A_1$ and $A_2$, and one two-dimensional
representation called $E$. In this respect it is similar to
Dashen's $\mathcal{Z}$-invariance discussed above. Hadrons
possess also internal symmetries, for baryons the internal symmetry
is usually
believed to be $SU(3)_c\times SU_{sf}(6)$. Hence, the geometric
wave functions (w.f.) must be combined with the internal w.f. in
such a way that the total w.f. are antisymmetric for baryons. The
spin-flavor group $SU_{sf}(6)$ has the representations referred to
as {\bf56}, {\bf70}, and {\bf20}. All baryons are commonly assumed
to fall down to the corresponding multiplets. Following the
w.f. antisymmetry principle, it was argued that $A_1$ must be combined with
{\bf56}, $A_2$ with {\bf20}, and $E$ with {\bf70}. Thus, the
states belonging to the representation {\bf70} are expected to be
parity doubled, while the states in {\bf56}
(they are known to include the ground states) should be parity
singlets. The geometrical considerations based on a baglike
analysis resulted in the claim that parity doubling in mesons
does not occur. Since at large $l$ the baryons and mesons become
similar, parity doubling in baryons should gradually disappear
as $l$ increases. The overall picture was in accord with the
available experimental data at that time.

Iachello's paper~\cite{ia} was followed by Robson's
comment~\cite{robson} and Iachello's reply~\cite{ia2}. The
discussion concerned a possibility for inclusion of the center of
Y-type string junction to the geometrical symmetries.

We would like to mention two instructive comments of current
importance which appeared in~\cite{ia,robson}. First, Robson~\cite{robson}
noted that the relativistic motion of the quarks and strings does
not allow a simple separation of total spin into orbital and
intrinsic spin components. The impact of relativity on such type
of models is difficult to assess. Second, Iachello~\cite{ia}
anticipated the failure of the nonrelativistic quark models in the
description of parity doubling. For instance, within
quark models with harmonic-oscillator potentials the states of
opposite parity have different numbers of oscillator quanta,
hence, parity doubling can be only accidental.

Another baryon string model explaining parity
doubling was proposed by Khokhlachev~\cite{kh}. The effect was
attributed to a large centrifugal potential for quarks in the
rotating gluon string. In this model two quarks are frozen at the
ends of linear gluon string and the third one moves along the
string. There are two levels with nearly equal energy
corresponding to "left" and "right" diquark states. These two
states can evolve into each other by quark tunneling under the
centrifugal barrier. The transition amplitude is small for large
$l$, hence, the mass difference of parity partners is small too.
An interesting prediction of the considered model is that the mass
difference dies off exponentially with increasing $l$,
\be
\Delta m_{\pm}\sim\sqrt{\frac{\mu}{L}}\exp(-\mu L),
\ee
where $L$ is the length of the string ($L^2=4(2l+1/2)/\pi\sigma$,
$\sigma$ is the string tension) and $\mu$ is the effective mass of
travelling quark when it moves at a large distance from the ends.

Independently, the available experimental information for meson and baryon Regge
trajectories of hadrons built of light quarks was summarized and
discussed by Kaidalov~\cite{kaidalov}. The data seemed to favor
the idea of approximate dynamical supersymmetry between mesons and
baryons (the related discussions have been occasionally appearing
in the literature,
see~\cite{anselmino} for a review). It was emphasized that the
existing quark models are unable to reproduce the observed
regularities in hadronic mass spectra, in particular, parity
doubling among baryons. The latter phenomenon was conjectured
to happen due to the Chiral Symmetry Restoration (CSR) for large
masses. It was noted also that CSR does not occur for the
principal boson Regge trajectories and that the behavior of boson
Regge trajectories can be explained by a smallness of spin-orbital
interaction between quark and antiquark. All these observations anticipated
qualitatively the main lines of later development of the subject
under consideration.

Nearly at the same time DeTar and Kunihiro
proposed~\cite{tk} a generalization of the linear sigma model
where two parity partner nucleons form a multiplet of the chiral
group and they can be degenerate with a non-vanishing mass. This model,
however, was intended to describe CSR at high temperatures
with entailing parity doubling of the baryon spectrum known from
the lattice simulations. But the idea itself was exploited later for
description of CSR in highly excited baryons.

\section{1990s: New ideas}

\subsection{Baryons in 1990s}

In 1990s the following idea independently came to mind of
different people: The systematic parity doubling in excited
baryons is nothing but a manifestation of effective chiral
symmetry restoration in the upper part of baryon spectrum. We have
just mentioned the idea of CSR in relation with Kaidalov's
work~\cite{kaidalov}. Kirchbach arrived at this idea in 1992
(see~\cite{kirchbach}) in a rather philosophical way, inspired by
an analogy with chirality in chemistry and biology. A manifest
realization of chiral symmetry above 1.3~GeV in non-strange baryons
was explicitly stated in~\cite{kirchbach2}. However, the systematic
occurrence of parity unpaired states and spin-parity clusters
forced her to abandon the idea of CSR in such a straightforward
interpretation and to propose an alternative scheme (to be discussed below).
Nevertheless the idea itself was not forgotten~\cite{riska,riska2}
(we refer to~\cite{kirchbach3} for relevant discussions). Nearly
at the same time CSR in excited baryons was independently observed
by J. Dey and M. Dey within a dynamical symmetry model (inspired
by 1960s Barut's work on dynamical conformal $O(4,2)$ group) based
on $U(15|30)$ graded Lie group reduced to $SU(3)$
subgroup~\cite{dey} (see also~\cite{dey2}). In framework of this
approach the baryons are supersymmetric partners of mesons.

After some years of recess the idea about different realization of chiral symmetry of
QCD in the low-energy and high-energy sectors (the Nambu-Goldstone mode
and the Wigner-Weyl one correspondingly) was again repeated
in the beginning of review~\cite{G1}, although the review
itself was devoted to the description of baryons within a
constituent quark model with the harmonic confinement potential.
In several years the potential models were criticized by Glozman~\cite{G2}:
They cannot explain the appearance of systematic parity doublets,
this is especially evident for the harmonic confinement. In
essence, the 10-years old Iachello's and Kaydalov's
conclusion~\cite{ia,kaidalov} was rediscovered. The
paper~\cite{G2} seems to be the first attempt to reveal the
dynamics underlying CSR.
The effect was ascribed to the strong
residual interactions between valence constituent quarks due to
Goldstone boson exchange. A parallel with the chiral phase
transition at high temperature was drawn. The proposed
explanation, however, did not work for mesons (the meson spectra
indeed did not exhibit parity doubling at that time).

To proceed further we should present the experimental spectrum for
nonstrange baryons, see Fig.~3 for nucleons and Fig.~4 for
$\Delta$-baryons. One can immediately notice the main features of
displayed spectrum --- parity doubling of many states and
clustering of masses near certain values of energy. The Particle
Data Group~\cite{pdg} averages the data over different experiments, this obscures
clustering because of accumulation of experimental errors. For
this reason it is instructive to demonstrate the results of a
separate comprehensive analysis. In Fig.~5 we show the data
provided by H\"{o}hler (this data is cited by the
Particle Data Group~\cite{pdg} under the name "Hoehler") for $\Delta$-baryons
(for nucleons the picture is very similar). Clustering in
Fig.~5 becomes much sharper. H\"{o}hler seems to be the first who
emphasized that baryons appear as spin-parity clusters rather than
as separate states~\cite{hol,hohl,hohl2}. Now these clusters often carry his name.
We draw attention to the (quasi)systematic parity singlets in Fig.~5
(or Fig.~4), especially the lowest $\frac32^+$, $\frac72^+$, and
$\frac{11}{2}^+$ states. One can expect that all states inside a
cluster are parity doubled except, in some cases, the state with
the highest spin. The existence of such parity unpaired
states represents a stumbling-block in interpretation of the parity
doubling phenomenon. Are they regular or we are simply dealing with
a lack of experimental data? At present this is not known, this
very point generates various models and speculations.

\begin{center}
\begin{figure*}
\vspace{-7cm}
\hspace{-3cm}
\includegraphics[scale=1]{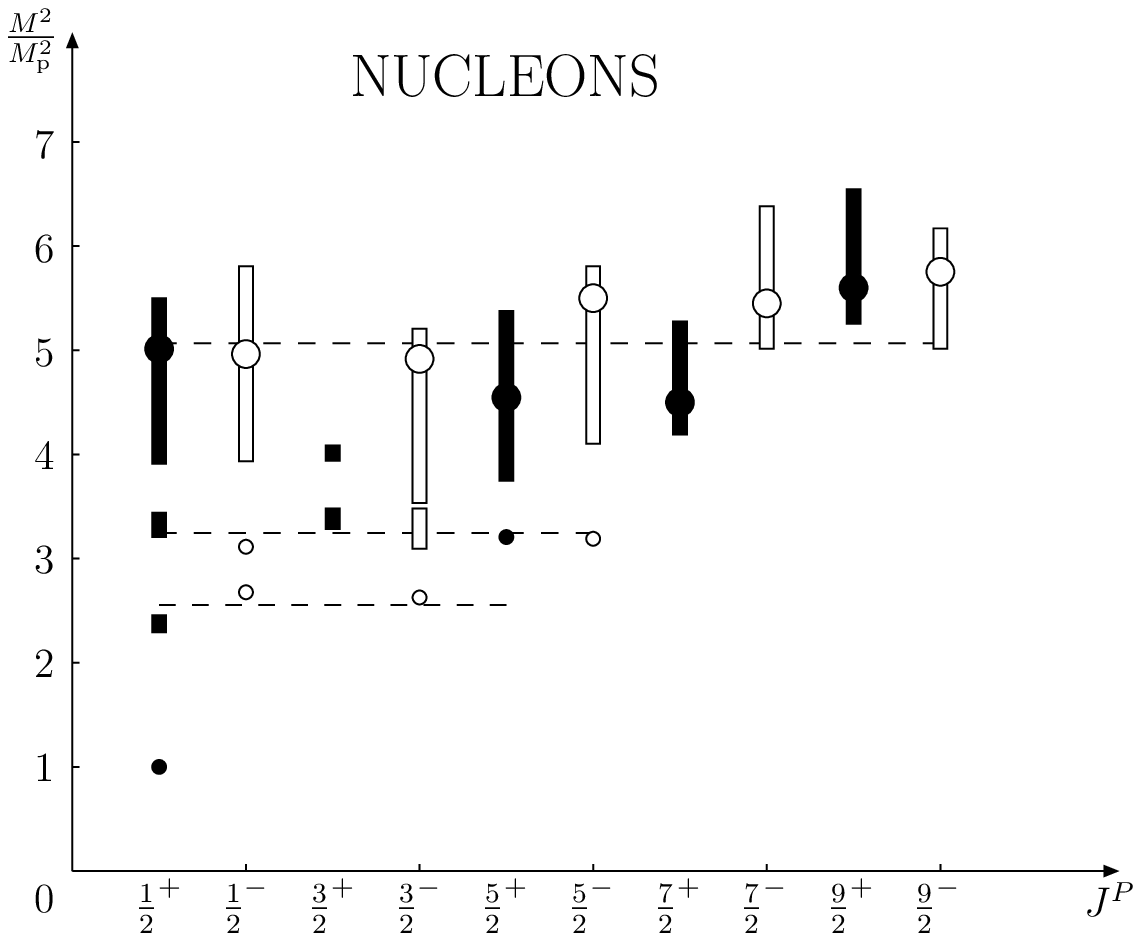}
\vspace{-16cm}
\caption{\small The experimental spectrum of nucleons~\cite{pdg} in units of the proton mass squared.
Experimental errors are indicated. The most reliable values
reported in~\cite{pdg} are denoted by circles. The filled and open
strips (circles) stay for the positive and negative parity states
correspondingly. The approximate positions of clustering are shown by dashed lines.}
\end{figure*}
\end{center}

\begin{center}
\begin{figure*}
\vspace{-6cm}
\hspace{-3cm}
\includegraphics[scale=1]{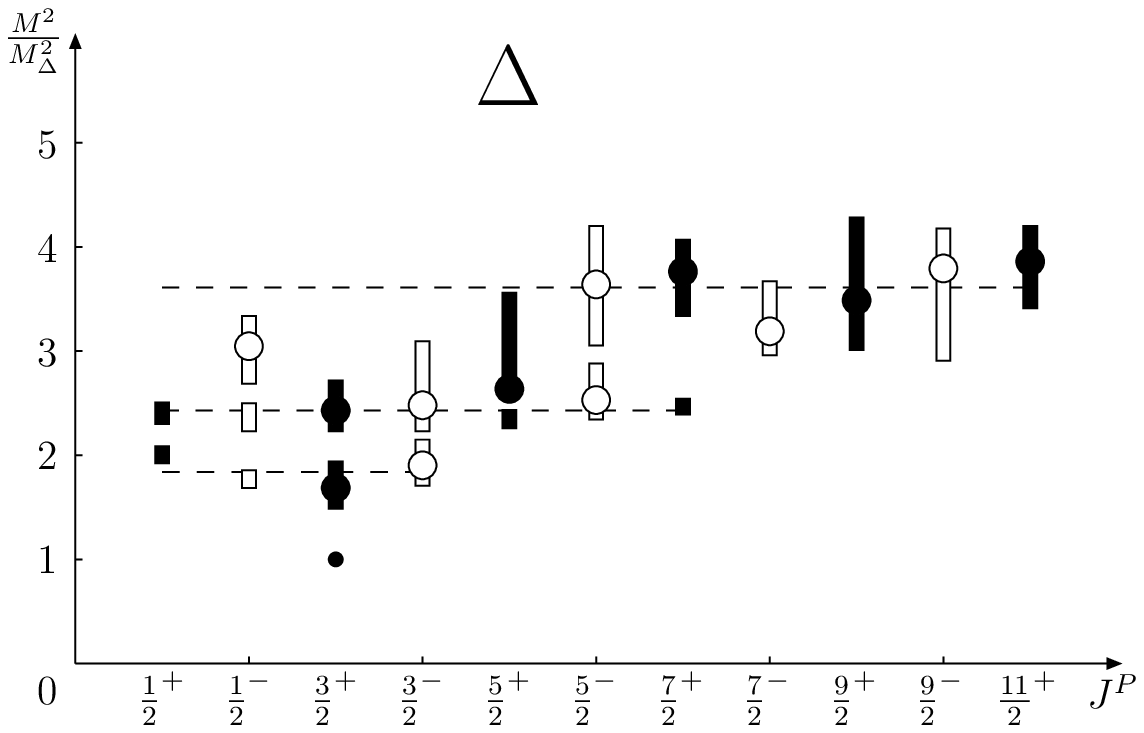}
\vspace{-18cm}
\caption{\small The experimental spectrum of $\Delta$-baryons~\cite{pdg} in units of the $\Delta(1232)$ mass squared.
The notations are as in Fig.~3.}
\end{figure*}
\end{center}

\begin{center}
\begin{figure*}
\vspace{-7cm}
\hspace{-3cm}
\includegraphics[scale=1]{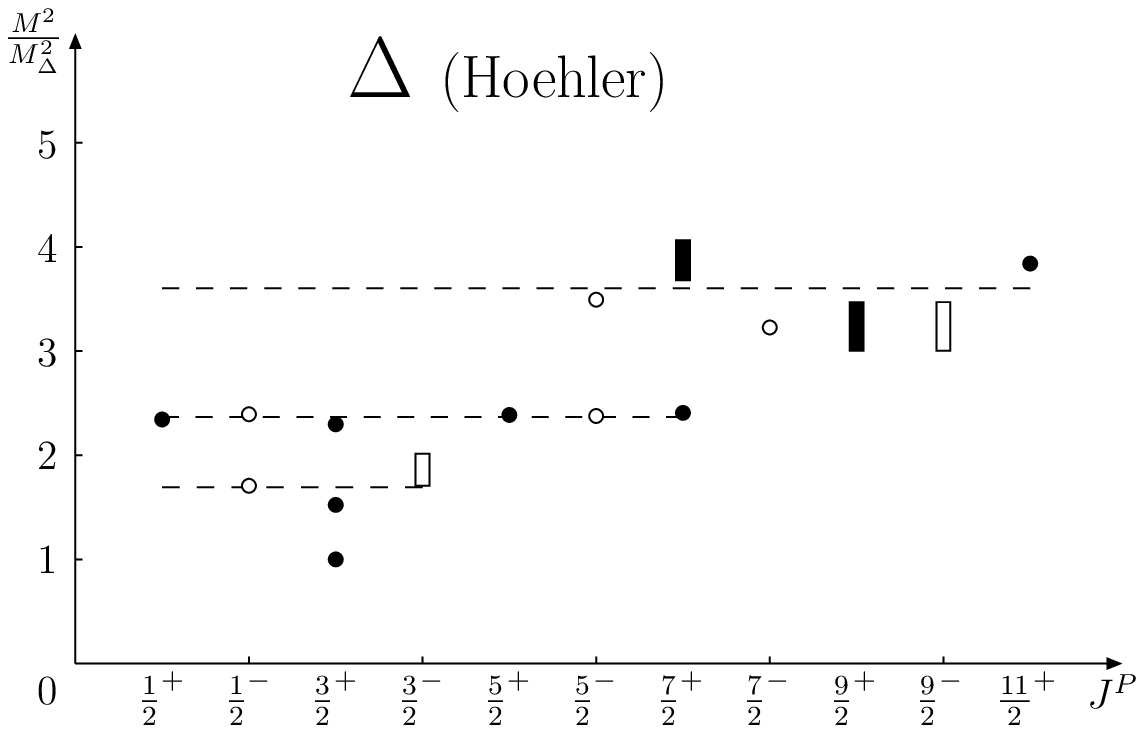}
\vspace{-18cm}
\caption{\small The spectrum of $\Delta$-baryons from H\"{o}hler analysis~\cite{pdg}.
The notations are as in Fig.~4.}
\end{figure*}
\end{center}

\vspace{-3.3cm}

The first theoretical explanation for H\"{o}hler's clusters was
proposed by Kirchbach~\cite{ki}. The symmetry of all reported nonstrange
baryon excitations was advocated to be governed by
$O(4)\times SU_I(2)$ rather than by $O(3)\times SU(6)_{sf}$
which is the usual textbook symmetry for classification of
the baryon states.
%The resulting picture of baryons somewhat resembles the theory of hydrogen atom.
The clusters appear due to the $O(4)$ partial
wave decomposition of the $\pi N$ amplitude, where only even valued
four-dimensional harmonics should be taken into account. In a
sense, it was a revival of old ideas of 1960's (see discussions
before and after Eqs.~\eqref{4h} and~\eqref{4h2}). These ideas,
however, were developed towards accommodation
of many new experimental data. The clusters of non-strange baryons
in Fig.~3 and Fig.~4 are assigned to $n=2,4,6$ poles of
$O(4)$ partial wave decomposition for the $\pi N$ amplitude. The
states inside each cluster fall into the Rarita-Schwinger-like
Lorentz multiplet (compare to Eq.~\eqref{alex}),
\be
\label{kirchM}
\left[\left(\frac12,0\right)\oplus\left(0,\frac12\right)\right]
\times\left[\frac{n-1}{2},\frac{n-1}{2}\right].
\ee
According to the proposed theory,
the states belonging to $n=2$ possess a natural parity $(-1)^l$,
the states in $n=4,6$ carry an unnatural parity $(-1)^{l+1}$. The
different assignment of parity is interpreted as appearance of
low-lying states on the top of
the scalar vacuum and that of high-lying states on the top of
the pseudoscalar vacuum (as the parity was defined as $\eta(-1)^l$,
with $\eta$ being the parity of underlying vacuum). The change of
underlying vacuum when passing to high excitations is suggested to
signal the chiral symmetry restoration in highly excited states. The
corresponding ideas and phenomenology were developed in~\cite{ki2,ki2b,ki2c,ki2d,ki2e,ki2g}.
In particular, a solution of the Velo-Zwanziger problem (the violation of
causality at propagation of the Rarita-Schwinger fields when minimally coupled
to an external electromagnetic field) was proposed: The low-spin
states entering the Rarita-Schwinger spinors should not be
eliminated as redundant components by some auxiliary conditions,
instead they should be treated as physically observable resonances
reflecting the composite character of baryons. Stated differently,
a pathology-free Rarita-Schwinger field describes a H\"{o}hler
cluster as a whole rather than a separate state (see~\cite{kinew,kinew2,kinew3,kinew4}
for the latest results). Kirchbach's classification
allowed to describe successfully the H\"ohler clusters and to reduce
significantly the number of "missing" states.

With regard to clustering in baryons we would like to make the
following remark. The first who predicted this phenomenon seems to
be Feynman. Basing on unpublished 1969 Caltech Lecture Notes, he
suggested certain approximate regularities among the square masses
of the baryons. His scheme was elaborated and published in~\cite{feynman}.
Now it appears to be timely to
remind the results. The proposed classification of baryons was guided
by the $SU_{sf}(6)$ quark model and the principle of Regge recurrence,
in other words, it was guided by certain "clustering" principle.
In the non-strange sector, a crucial test for the suggested mass
degeneracies had to be the discovery of six "missing" states. In
addition, the confirmation of these states was claimed to be
equivalent to "{\it... the statement that the spin-orbit contribution to
the mass splitting in the quark model is small}".
This guess-work, likely, was not taken seriously by specialists (at least,
Feynman {\it et al's.} paper~\cite{feynman} has an extremely low
citation by non-authors, which is quite unusual for such a physicist as Richard
Feynman). Curiously enough, later all these
six "missing" states were gradually discovered with the masses
close to Feynman's predictions! At present~\cite{pdg} they are
(we display the star rating):
$N_{\frac32^+}(1720)^{****}$,
$N_{\frac32^-}(1750)^{***}$,
$\Delta_{\frac32^+}(1920)^{***}$,
$N_{\frac32^-}(2080)^{**}$,
$N_{\frac52^-}(2200)^{**}$, and
$N_{\frac92^-}(2250)^{****}$.
Unlike H\"{o}hler's spin-parity clusters, Feynman's clusters are
only "spin" ones, they do not predict parity doubling.

Another approach to the problem of baryon parity
doublets was suggested by Balachandran and Vaidya~\cite{bv,bv1}.
They noticed that parity doublets occur
typically in systems with two differing time scales. There are
numerous examples of this phenomenon in molecular and nuclear
physics. The possible parity doubles in particle physics were
supposed to have the same origin. The idea was then realized
in~\cite{bv2}, where the baryon was modeled by slow Skyrmion and
fast light quarks whizzing around.

\subsection{Mesons in 1990s}

In 1990s there was increasing evidence that meson states of different
spin fall into degenerate towers at a given mass, this interesting
tendency attracted some attention within the framework of
relativistic quark models~\cite{goebel}, although the related
problem of parity doubling was not directly addressed.
The experimental data did not
unambiguously show a systematic parity doubling among mesons.
As a consequence, physicists were not enthusiastic to work
in the given direction. We are aware of one attempt to address the
problem directly, within a combined analysis of effective quark models and
asymptotic sum rules from QCD. Before the relevant discussions we remind
some prehistory of asymptotic sum rules.

In early 1960s the idea of asymptotic chiral symmetry
appeared\footnote{The roots of this idea go back to 1950s when different
authors were attempting to uncover a "higher symmetry" of strong
interactions, which is broken at low energies but perhaps becomes
exact in some high energy limit (see, e.g.,~\cite{coleman} for
references).}.
This symmetry was supposed to become rigorous at sufficiently
high energy region where the symmetry breaking effects are
negligible\footnote{In the case of badly broken symmetries, it is
necessary to indicate the limit, where the symmetry is present in
pure form. By the beginning of 1960s it became clear that the
relevant limit is the limit of high frequencies, i.e. of high
energies~\cite{zach}.}.
For instance, the axial nucleon current $j_{\mu}^A$ is
not conserved by itself, the Partially Conserved Axial Current
(PCAC) hypothesis states that
\be
j_{\mu}^A\sim\bar{\psi}_N(i\gamma_{\mu}\gamma_5+2m_N\gamma_5p_{\mu}/p^2)\psi_N,
\ee
where the second term is associated with the pion and $m_N$ is the nucleon mass.
However, if the momentum is so large that $p^2\gg m^2_N$ then one
does not need PCAC, the axial current is conserved by itself. This
is a reflection of the fact that the kinetic term
$\bar{\psi}_N\gamma_{\mu}\partial_{\mu}\psi_N$ in effective
strong-interaction Lagrangians becomes dominant in the high-energy
region. Such a point of view was often stressed by
Nambu~\cite{nambu1964} (see also~\cite{nambu1960}).
Consider as an example the $\pi N$ system.
The corresponding amplitude possesses a broken chiral
invariance, the chirality is conserved due to pions. However, if
the momentum is so large that the nucleon mass may be neglected,
one does not need the pions to conserve chirality. The $\pi N$
scattering amplitude becomes chirally invariant
by itself, hence, the soft pion emission process will vanish.
This observation results in interesting predictions~\cite{nambu1964}.

The Weinberg's sum rules~\cite{wein} are, perhaps, the most famous application of
the asymptotic chiral symmetry concept. Consider a two-point
correlation function for hadron currents (for the sake of convenience
we consider the momentum representation in Euclidean space),
\be
\Pi_k(Q^2)\sim\int\! d^4\!x\, e^{iQx}\langle j_k(0)j_k(x)\rangle,
\ee
where $k$ denotes a set of indices characterizing the hadron current $j_k(x)$.
Let $j_+(x)$ and $j_-(x)$ be parity (chiral) partner currents. Consider the
difference of their two-point correlators and impose the condition
\be
\label{ps1}
\Delta(Q^2)=\Pi_+(Q^2)-\Pi_-(Q^2)\xrightarrow[Q^2\rightarrow\infty]{}0,
\ee
This condition can be regarded as a mathematical expression for the
asymptotic chiral symmetry~\cite{sakurai}. It gives relations for the hadron
masses when one makes use of the pole approximation. Weinberg
considered the vector and axial-vector isovector currents, and
assumed the dominance of the ground state in the pole
approximation and the convergence condition
\be
\label{ps2}
Q^4\Delta^t(Q^2)\xrightarrow[Q^2\rightarrow\infty]{}0,
\ee
where $\Delta^t$ means that the transverse part
$(-\delta_{\mu\nu}Q^2+Q_{\mu}Q_{\nu})$ is factorized out.
Taking into account PCAC and the so-called KSFR relation
($Z_{\rho}=2f_{\pi}^2$, here $Z_{\rho}$ is the $\rho$-meson residue and
$f_{\pi}\approx93$ MeV is the weak pion decay constant), he derived the relation
$m_{a_1}^2=2m_{\rho}^2$, which was in impressive agreement with
the experimental data at that time. The idea turned out to be very
fruitful. For instance, very soon Das {\it et al.}~\cite{das} calculated the pion
electromagnetic mass difference due to the Weinberg sum rules.
Schechter and Venturi~\cite{venturi} showed that under some
assumptions the Weinberg relation
can be reproduced numerically from the values of neutron and proton magnetic
moments and the axial constant. One of their assumptions was that
the baryons can be assigned to a definite chiral representation at
very large momentum transfer (compared to the baryon masses), i.e.
again one used the asymptotic chiral symmetry.
The concept of asymptotic symmetries became a standard topic in
the textbooks on elementary particles of that time (see,
e.g.,~\cite{sakurai}).
Later Weinberg's assumptions were somewhat justified. The pole approximation,
i.e. the approximation of infinitely narrow resonances, is
equivalent to the large-$N_c$ limit in QCD~\cite{hoof,hoofw}. The
convergence condition~\eqref{ps2} was derived within the Operator
Product Expansion (OPE) method~\cite{svz,svz2}.

Consider the exact planar limit of QCD (infinite number of
colours). As a rule, this limit is known to work well within 10\% accuracy.
The meson correlators then have to be saturated completely
by the infinitely narrow meson resonances~\cite{hoof,hoofw}. The number
of resonances with identical quantum numbers should be infinite in order to reproduce the
perturbative logarithmic asymptotics of correlators.
Thus, one has
\be
\label{ps3}
\Delta_{\text{planar}}(Q^2)=\sum_{n=0}^{\infty}\frac{Z_+(n)}{Q^2+m^2_+(n)}-
\sum_{n=0}^{\infty}\frac{Z_-(n)}{Q^2+m^2_-(n)}\xrightarrow[Q^2\rightarrow\infty]{}0.
\ee
Here $n$ is analogous to the radial quantum number.
The OPE~\cite{svz,svz2} predicts a quite rapid convergence at large
Euclidean momentum in the r.h.s. of Eq.~\eqref{ps3} (say, as
$\mathcal{O}(Q^{-4})$ for the scalar case and as $\mathcal{O}(Q^{-6})$
for the vector one).
On the other hand, the dominance of ground state
($n=0$) is typically a good approximation. In order to reconcile
these facts one can deduce that the masses and residues of
opposite-parity states should be rapidly converging with $n$.
A similar reasoning forced A. A. Andrianov and V. A. Andrianov~\cite{aa} to
conclude that a rapid restoration of chiral symmetry in Eq.~\eqref{ps3}
suggests a rapid CSR in the spectrum of radially excited mesons.
Consequently, any effective quark model describing the strong dynamics
above the chiral symmetry breaking scale ($\approx1$ GeV) has to
reproduce the asymptotic restriction~\eqref{ps3} dictated by OPE,
i.e. it has to reproduce the CSR at high energies. This is a
powerful test for QCD-motivated effective quark models even if they do not
describe the radial excitations. The corresponding concept was
formulated earlier~\cite{AV}. Later,
matching of some effective models to the short distance behavior
of two-point correlators was performed~\cite{other,other2,other3,other4,other5}.
In addition, since chiral symmetry is restored quite rapidly, already the first
radial excitation might reveal this phenomenon, i.e. one should
have then $m_+(1)-m_-(1)\ll m_+(0)-m_-(0)$. This property was
demonstrated for the case of so-called quasilocal quark
model in the scalar channel~\cite{aa}. If one assigns the first
scalar and pseudoscalar "radial" excitations to the states
$f_0(1370)$ and $\pi(1300)$ then this prediction is fulfilled indeed.
Moreover, a fast CSR in the spectrum was
argued~\cite{aa} to entail the decoupling of heavy parity doublets
from the low-energy pion physics. In practice, this statement
means that contribution of radial excitations to the constants of
low-energy effective chiral Lagrangians~\cite{gasser} is negligible, these
constants are mostly saturated by the ground states.

\section{2000s: Golden age}

\subsection{General discussions}

The beginning of this decade is marked by an experimental
breakthrough in the unflavored meson sector. The analysis
of Crystal Barrel Collaboration experimental data on proton-antiproton
annihilation in flight in the energy range 1.9-2.4 GeV revealed
more than thirty new resonances (see, e.g.,~\cite{anietall,anietall2,anietall3,anietall4}).
Subsequently, all known light mesons
were systematized by Anisovich with collaborators
in~\cite{ani,ani2}, which resulted in the experimental discovery
of approximately linear trajectories on the $(n,M^2)$ and
$(J,M^2)$ planes ($n$ is the "radial" quantum number and $J$ is the
meson spin). In particular, on the $(n,M^2)$ plane the light mesons
can be fitted with a good accuracy by the linear parametrization:
\begin{equation}
\label{lin}
M^2(n)=m_0^2+an, \qquad n=0,1,2,\dots,
\end{equation}
where $m_0$ is the mass of basic meson and $a$ is the
trajectory slope parameter. The latter turned out to be
approximately the same for all trajectories,
$a=1.25\pm0.15$~GeV$^2$. It is exactly a string-like
spectrum predicted by many dual models and effective boson string models
starting since 1960s.
However, since these experimental
results were extracted by a single group, many of them are still listed by
Particle Data~\cite{pdg} as not well confirmed states. The latest
review on the Crystal Barrel results is contained in Bugg's
work~\cite{bugg}, the averaged slope of meson trajectories was
reported there to be\footnote{This value coincides with the fit
obtained independently in~\cite{afonrec} and is very close to an
earlier estimation~\cite{pancheri} which did not make use of the
Crystal Barrel data. A minor discrepancy between Anisovich's and
Bugg's fits originates from a different treatment of the scalar
sector.} $a=1.14\pm0.013$~GeV$^2$.

The analysis of Crystal Barrel data was criticized for adopting
resonances from the outset, this interpretation of data, however,
is in accord with the general principles of quantum field theory,
such as analyticity. This point seems to be underestimated by
other groups who have the data. It should be mentioned that the
analysis was performed without any intentions to obtain something
like linearity. It is quite remarkable that the final
systematization of the best fits yielded (unexpectedly!) the
linear trajectories and numerous cases of parity doubling.

We display some typical examples of meson trajectories from~\cite{ani2} in
Fig.~6.
A prominent feature of presented plots is duplication of some
trajectories. This effect is trivial for the scalar mesons: The
lower trajectory corresponds to the states in which the strange
component dominates. In other cases the explanation can be given in
terms of nonrelativistic quantum mechanics~\cite{ani}. Let $\vec{l}$ and
$\vec{s}$ be relative angular momentum of
quark-antiquark pair and its spin. The $P$ and $C$
parities are defined for quark-antiquark pair as $P=(-1)^{l+1}$ and
$C=(-1)^{l+s}$. Following the rule for the total spin
$\vec{J}=\vec{l}+\vec{s}$,
the vector $IJ^{PC}$ state $11^{--}$ can be
constructed by two ways, $(l,s)=(0,1),(2,1)$ (the S- and D-wave
vector mesons in usual spectroscopic language), while its parity
partner $11^{++}$ is made by one way only, $(l,s)=(1,1)$. The same can be
repeated for all other cases. Thus, at $J>0$ one of parity
conjugated radial trajectories is always duplicated.

\begin{center}
\begin{figure*}
\vspace{-7cm}
\hspace{-3cm}
\includegraphics[scale=1]{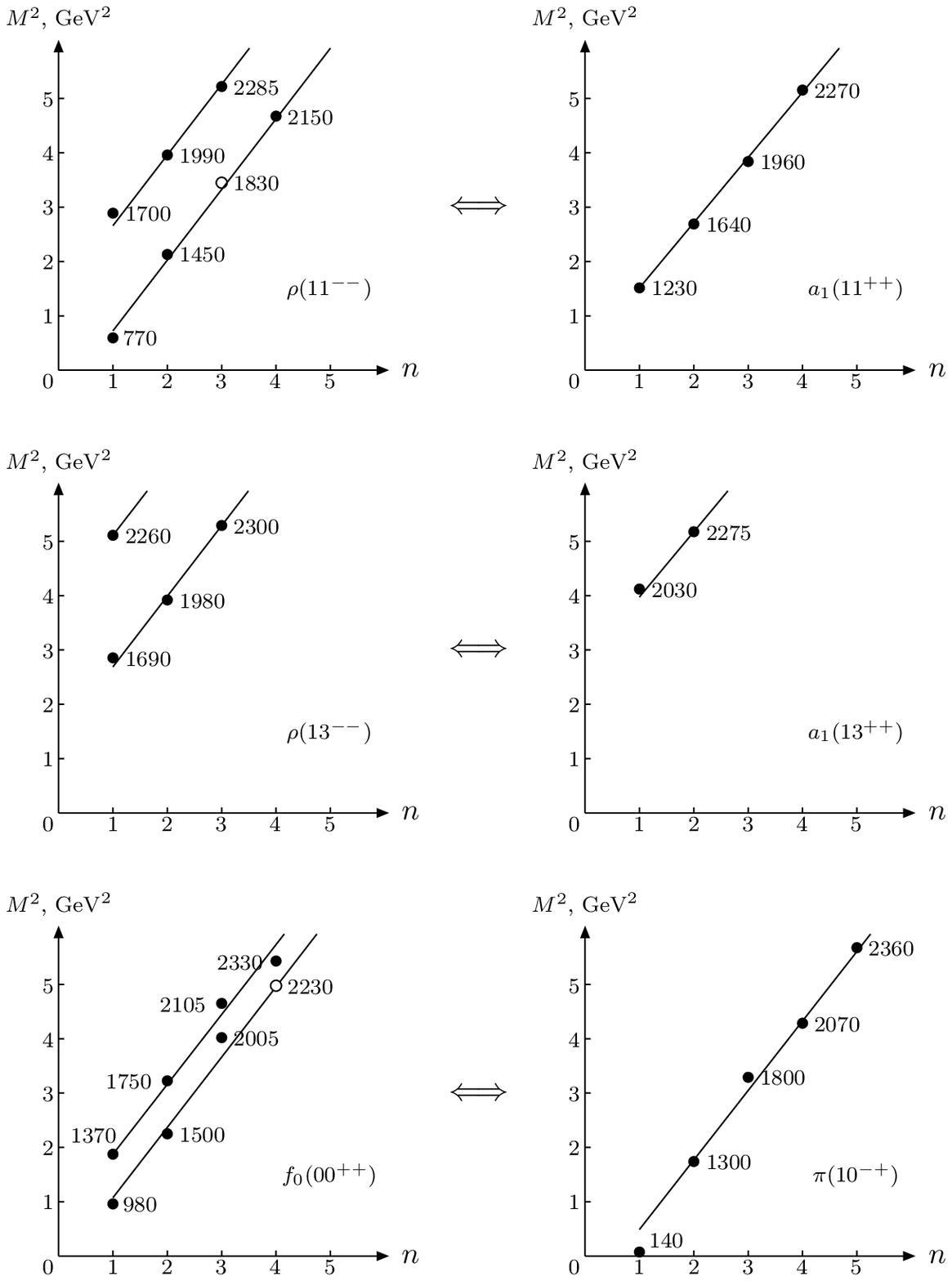}
\vspace{-8cm}
\caption{\small Some plots for the $IJ^{PC}$ spectrum of unflavored mesons from~\cite{ani2}.
Open circles stay for the predicted states. Chiral partners are marked by "$\Leftrightarrow$".}
\end{figure*}
\end{center}

\vspace{-1.3cm}

%However, the classification in terms of $n^{2s+1}l_J$ $\bar qq$
%states employed in~\cite{ani} looks natural for the systems made
%of heavy quarks, where the potential models may be successfully
%applied. In the light quark sector the relativistic effects are
%supposed to be drastic. Consequently, the description based on the
%nonrelativistic quantum mechanics seems to be unjustified for this
%sector. Although the given classification turned out to be quite
%successful, it is important to explain the spectrum from the first principles or

At the beginning only a little attention was paid to the Crystal
Barrel results. Many specialists somehow overlooked them. The
baryon sector was remaining richer, hence, more interesting.

Jido {\it et al.}~\cite{jhk} basing on ideas of DeTar and Kunihiro~\cite{tk}
proposed to organize low-lying baryon
fields into the representations
$(1,\frac12)\oplus(\frac12,1)$ of $SU(2)_L\times SU(2)_R$
chiral group (the so-called quartet scheme or mirror assignment).
Soon Cohen and Glozman~\cite{G23}, basing on the quark-hadron
duality,
advocated the point of view that the quartet scheme
should be applied to physical states of highly-lying
baryons (in~\cite{jhk} it was applied to low-lying baryon fields,
the physical states are obtained after acting by fields on the
vacuum, the difference would not be critical were the chiral
symmetry unbroken in the physical vacuum), the CSR then ensues.
In addition, CSR was substantiated by an OPE-based reasoning
applied to the case of baryon currents. This reasoning is somewhat
similar to that of described at the end of previous subsection for mesons.
In this respect the baryon case, however, is more involved as the baryonic
excitations are not well separated even in the planar limit, where
they emerge as solitons forming rather a continuum~\cite{hoof,hoofw}.

We explain briefly the $SU(2)_L\times SU(2)_R$ classification of
baryons performed in~\cite{G23}. A general irreducible
representation is marked by two indices $(I_1,I_2)$. It transforms
under parity into $(I_2,I_1)$. As long as QCD is invariant under
parity reversal, one cannot ascribe any definite parity to the
states in $(I_1,I_2)$ unless $I_1=I_2$. Hence, if chiral symmetry
is restored the multiplets must either be $(I,I)$ or
$(I_1,I_2)\oplus(I_2,I_1)$. The latter is an irreducible
representation of the parity-chiral group\footnote{Such a group-theoretical
machinery is familiar from the representation theory for the groups where
representations are formed in terms of the sum of two objects,
and a parity reversal changes the sign of one, but not the other.
A classical example is the Lorentz group, whose representations are
formed by the sum of rotation generators (pseudovector) and boost generators
(vector). An irreducible representation $(j_1,j_2)$ of the
restricted Lorentz group (i.e. without time and parity reversal)
is carried into the $(j_2,j_1)$ representation by a parity
reversal. Thus it is not a representation any more for the full
Lorentz group, unless $j_1=j_2$. The irreducible representations
of the latter are given by $(j_1,j_2)\oplus(j_2,j_1)$. As a
consequence, the relativistic field Lagrangians respecting
parity should not contain the left- and right-handed
Weyl spinors because they are Lorentz
$(1/2,0)$ and $(0,1/2)$, one must use Dirac spinors
$(1/2,0)\oplus(0,1/2)$, the vector potentials carry
$(1/2,1/2)$, while the field tensors transform under
$(1,0)\oplus(0,1)$, the $(1,1/2)\oplus(1/2,1)$ is the
Rarita-Schwinger field representation, etc.},
which is isomorphic to $O(4)$. As baryons in two flavor QCD
have half integral isospin, they cannot be in $(I,I)$
(since the isospin {\bf I} of states in $(I_1,I_2)$ is
$|I_1-I_2|\leq\text{\bf I}\leq I_1+I_2$).
Thus, if chiral symmetry is effectively restored, the baryons fall
into $(I_1,I_2)\oplus(I_2,I_1)$ parity-chiral multiplets with
$I_1$ integral and $I_2$ half integral. The simplest
representations are: i) $(1/2,0)\oplus(0,1/2)$ (nucleons),
ii) $(3/2,0)\oplus(0,3/2)$ (deltas), iii) $(1,1/2)\oplus(1/2,1)$
(the nucleons and deltas in one multiplet --- the mirror
assignment). Consider baryons with a fixed spin. If the cases i)
and ii) are realized then the mass of parity doublets in nucleon
spectrum {\it a priori} do not coincide with that of in delta one
because these dublets belong to different representations. In case
iii) the masses of nucleon and delta parity dublets coincide. The
phenomenological analysis favored the quartet realization iii)
in the experimental spectrum.

The upshot is that the effective CSR is not simply equivalent to
parity doubling (such an equivalence, however, takes place for the effective
axial $U(1)_A$ symmetry restoration), it manifests itself through
the existence of complicated multiplet structure. Originally this point was
accentuated by Jaffe~\cite{jwork}.

Subsequently Glozman {\it et al.} elaborated the ideology of effective
restoration of chiral and axial symmetries, having already written
about two tens papers on the subject. A comprehensive review of
this line of research is contained in~\cite{glozrev}, we refer to
this review for a detailed reading.

We will discuss only some aspects described in~\cite{glozrev}. The
parity-chiral classification was applied to mesons, the idea was
partly borrowed from Cohen and Ji work~\cite{CJ}, where possible
chiral representations for hadron interpolating currents were
systematically classified\footnote{The first who suggested the
parity-chiral assignment to known and predicted mesons (within
more general $SU(3)_L\times SU(3)_R$ chiral group) seems to be
Arnold~\cite{arnold}. This proposal was essentially based on
exchange degeneracy introduced in the same paper. The problem was
posed to relate dynamics to the suggested symmetry and in this way
to derive the pattern of mass splitting inside multiplets. The
general solution of this problem remains to be a dream up to now.}.
This allowed to explain the duplication
of some trajectories in Fig.~6 without use of nonrelativistic
notions. For instance, the reason for duplication of $\rho$-meson
trajectory is that there are two kinds of $\rho$-meson falling
into different representations of parity-chiral group\footnote{The
existence of vector mesons differing by chiral properties has been
proposed earlier, we refer to~\cite{sh} for relevant discussions.}.
The $\rho$-mesons of the first kind transform as
$(1,0)\oplus(0,1)$, their chiral partners are the $a_1$-mesons.
The interpolating currents are $\bar{q}\vec{\tau}\gamma_{\mu}q$
and $\bar{q}\vec{\tau}\gamma_{\mu}\gamma_5q$ respectively. The
$\rho$-mesons of the second kind carry $(1/2,1/2)$ representation,
their chiral partners are the $h_1$-mesons which are isosinglets.
The interpolating currents are $\bar{q}\vec{\tau}\sigma_{\alpha\beta}q$
(or $\bar{q}\vec{\tau}\partial_{\mu}q$) and
$\bar{q}\varepsilon_{\alpha\beta\gamma}\sigma_{\beta\gamma}q$
(or $\bar{q}\gamma_5\partial_{\mu}q$) respectively. Thus, the
underlying reason for duplication of vector mesons is that they have
two different interpolating currents belonging to different
representations of parity-chiral group.
The naive interpretation of CSR as simply parity doubling would
lead to inconsistency: Experimentally there exist about two times more excited
vector states than the axial-vector ones.
At low energies the multiplets are mixed and $\rho(770)$ meson is roughly an equal
mixture of both representations. If the chiral symmetry is
restored at high energies this mixing disappears. Then the
combined amount of highly excited $a_1$ and $h_1$ mesons must
coincide with the amount of $\rho$-meson excitations.
This situation recurs
for higher spins. The overall scheme enjoined a phenomenological
success in description of the Crystal Barrel data on unflavored
mesons~\cite{ani2,bugg}.

The justification for the effective restoration of chiral and axial
symmetries of two flavor QCD Lagrangian in the spectrum of highly
excited states relies heavily on quasiclassical considerations.
It was advocated that quantum fluctuations are suppressed at
energies large enough, hence, the highly excited systems
necessarily become quasiclassical. As a consequence, the classical
symmetries of QCD should be effectively restored.
Various examples are cited in~\cite{glozrev} showing that in high
excitations the low-energy effects become inessential and the
broken symmetries are restored high in the spectrum.
As a matter of fact, in the quantum theory this property is well known
--- properties of any quantum system approach to its classical ones while the
quantum numbers defining the stationary states of this system are
large enough (see, e.g., a standard textbook~\cite{landau}). For
instance, the wave function corresponding to the $n$-th radial
excitation has approximately $n$ nodes in the coordinate space,
the distance between two
neighboring nodes is of order of de Broglie wavelength, hence, at
large $n$ one inevitably obtains a quasiclassics. The assumption
is that one has something similar in QCD in the resonance region.

The semiclassical expansion suggests that the higher Fock
components in the quark wave functions are suppressed for high
radial and orbital excitations. In this situation, the string
picture with massless relativistic quark and antiquark at the ends seems to
be an admissible approximation for non-strange mesons. An idea was
put forward that CSR then follows if relativistic quark and
antiquark at the ends have definite chiralities. In this picture
any degenerate chiral pair belongs to the same intrinsic quantum
state of the string, the opposite parity of states in the pair
results from different chiral configurations of the quarks at the
ends. It was noted also that the string picture should lead to a
much higher degeneracy than just parity doubling. This issue will
be discussed in Section~7.

In the recent review by Jaffe {\it et al.}~\cite{jaffenew}, three candidates for
possible origin of parity doubling among non-strange baryons have been finally
selected. The first possibility is a dynamical suppression of the violation of
flavor singlet axial symmetry of QCD. This scenario is substantially different
from $U(1)_A$ restoration (i.e. an effective Wigner-Weyl realization of
axial symmetry in highly excited baryons) and can be realized if the matrix
elements $\tilde{G}^{\mu\nu}G_{\mu\nu}$ ($G_{\mu\nu}$ is the gluon field tensor)
taken between baryons of opposite parity are very small for some reasons.

The second possibility is that parity doubling in the baryon spectrum might be
related to the internal geometrical structure of baryons ("deformed shapes")
by analogy with
similar phenomena in nuclear and molecular physics. We have already discussed
such possibilities in relation with Iachello's~\cite{ia} and Balachandran
and Vaidya's~\cite{bv,bv1,bv2} models. The underlying observation is that if
an intrinsic state (whose collective quantization describes a system in question)
spontaneously violates parity and the deformation of this state is relatively
rigid, then the low-lying excitations of the system will display parity doubling.
The term "rigid" means that the Hamiltonian matrix element between the intrinsic
state and its parity image is small. In case of baryons this intrinsic state
might be an elongated quark-diquark structure which violates reflection symmetry.
If tunneling of a "mobile" quark from one end to the other were effectively suppressed,
then parity doubling would result. Khokhlachev's model~\cite{kh} discussed
above is nothing but a particular realization of this idea.

The third possibility consists in accommodation of parity doubling within models which
do not naturally lead to this phenomenon. Weinberg's "mended symmetry approach"~\cite{wein2,wein3}
was considered as an example. The underlying hypothesis of this picture is that
chiral symmetry may be realized on the matrix of the couplings of the Goldstone
bosons rather than on the mass eigenstates. The mass matrix
$\hat{m}^2$ at any given helicity may be then written as the sum of a chiral scalar $\hat{m}_0^2$ and the fourth
component of a chiral four-vector $\hat{m}_4^2$ with respect to $SU(2)_L\times SU(2)_R$ formed
by the isospin and the axial vector coupling matrix.
The term $\hat{m}_4^2$ appears due to existence of vacuum Regge trajectories and
destroys the algebraic chiral symmetry for the eigenstates. To obtain a model of hadron
one should choose a reducible representation of this $SU(2)_L\times SU(2)_R$ and mixing angles.
Parity doubling emerges for some particular choices of representation contents and corresponding
mixing angles.

Klempt~\cite{kl} proposed that the
appearance of parity doublets in light baryons does not reflect the chiral symmetry
but rather the vanishing of spin-orbit forces ($\vec{l}\cdot\vec{s}$) in quark
interactions; the chiral symmetry itself might lead only to a weak
attraction between chiral partners. The idea was somewhat inspired by
Feynman {\it et al's.} analysis~\cite{feynman} of baryon mass regularities mentioned above.
In comparison with the
hypothesis of CSR, this interpretation gives different
predictions for the spectrum of highly excited baryon resonances,
grouping them into $(l,s)$ multiplets. Baryons emerge as
approximately mass degenerate clusters where
both parity doublets and parity singlets can live. In addition, baryons
can be assigned to $(l,s)$ multiplets so that the linear mass formula for baryons holds~\cite{kl2}
\be
\label{lin2}
m^2=m_{0,k}^2+a(l+n_r),
\ee
where the intercept $m_{0,k}^2$ depends on the type of baryons (labelled by $k$),
the slope $a$ is the slope of principal meson Regge trajectories
(compare to Eq.~\eqref{lin}), and $n_r$ is the "radial" quantum
number. Taking into account the nonrelativistic definition of parity for baryons,
$P=(-1)^l$, Eq.~\eqref{lin2} yields a pattern of parity doubling among baryons.

Finally, we mention a recent new scheme for parity doubling among
light mesons based on the MacDowell symmetry~\cite{afonrec}. The ides
is that if a kind of dynamical meson-baryon supersymmetry exists
indeed, then  the MacDowell symmetry among the baryons should have
an imprint on the meson spectrum. The phenomenological analysis
carried out in~\cite{afonrec} using the Crystal Barrel
data~\cite{bugg} showed that this proposal looks really plausible,
at least formally.

\subsection{Parity doubling in effective quark models}

Under "effective quark models" we mean the following. Let us
imagine that we have "integrated" over all gluons
and over other degrees of freedom above some energy scale in the QCD Lagrangian.
The obtained Lagrangian should then describe the strong
interactions below the chosen scale, with the result of "integration"
being encoded in types of residual interactions and in values of
coupling constants (in some models the "integration" over all fermion degrees of freedom
is also assumed). As long as nobody knows how to perform this
analytically, one often resorts to models in studying the
low-energy dynamics of strong interactions. Any such model puts forward
an effective Lagrangian (or Hamiltonian) as an interpolating Lagrangian for
the "genuine" effective Lagrangian of QCD. The better the relevant
effective degrees of freedom are guessed, the better is the effective model.

In most cases parity doubling and CSR within effective quark models have been
studied within various extensions of the Nambu--Jona-Lasinio (NJL) model~\cite{njl,njl2}.
The NJL model approximates the low-energy strong interaction dynamics by a local four-fermion
interaction, the corresponding Lagrangian density can be written as
\be
\label{njllag}
L_{\text{NJL}}(x)=\bar{q}(x)(i\!\!\not\!\partial-m)q(x)+g_ij_i(x)j_i(x),
\qquad  j_i(x)=\bar{q}(x)\Gamma_iq(x).
\ee
where $\Gamma_i$ is a Lorentz and isospin structure fixing the quantum numbers of interpolating
current $j_i(x)$ (the scalar one in the original NJL)
and the summation over $i$ is assumed. To specify
a model completely one needs to fix a calculation method and
momentum cutoff (of the order of the chiral symmetry breaking scale, 1 GeV, in conventional NJL).
Originally the idea came from solid state physics where such Lagrangians are used
for phenomenological description of superconductivity\footnote{The BCS theory is meant.
The situation with superconductivity is similar --- nobody knows how
to derive the BCS Lagrangian from the underlying Quantum Electrodynamics.}.
In particle physics it was first applied to nucleons and later,
after advent of QCD, to quarks.
The Lagrangian~\eqref{njllag} can be easily
made chiral-invariant, e.g. for $j_1(x)=i\bar{q}(x)q(x)$, $j_2(x)=\bar{q}(x)\vec{\tau}\gamma_5q(x)$,
and $g_1=g_2$ one has a chiral-invariant scalar-pseudoscalar effective model.
The scalar four-fermion interaction dynamically breaks the chiral invariance
if the value of coupling constant $g_i$ exceeds some critical
value --- the corresponding mass-gap equation reveals a non-zero solution for the
vacuum average $\langle\bar{q}q\rangle$. The overall effective
theory has a substantially right low-energy physics as
practically all important relations
of current algebra can be naturally reproduced within the NJL
model. This model describes commonly the ground boson states,
although sometimes it is used for baryons as well.

The number of proposed extensions for the NJL model is so large
that even all extensive reviews on this model taken together
reflect only a small fraction of related researches. Loosely speaking,
we would classify the proposed extensions by means of (or mixing of)
four directions. The first
way is to incorporate the quark currents with new quantum numbers
with the aim to describe the hadrons possessing these new quantum
numbers. The second direction is to include higher-dimensional
vertices. For instance, to take into account the $U(1)_A$ symmetry breaking
one needs the six-fermion 't Hooft term. These two extensions,
however, describe generically the ground states only. As long as
parity doubling is expected in high excitations, one needs something
qualitatively different. The third and fourth types of extensions
have been developed for accommodation of higher excitations.

The third possibility is to consider nonlocal interactions,
\be
L_{\text{int}}(x)\sim g_i\int V_i(x-y)j_i(x)j_i(y)d^4\!y.
\ee
The functions $V_i(x-y)$ are called formfactors or potentials
(the latter name is inherent in the Hamiltonian formulation when
one solves the Bethe-Salpeter equation for bound states, the term "kernel"
is also used). The choice of these functions specifies a model
(see, e.g.,~\cite{volkov,volkovr}). To our
knowledge, the first successful application of such an extension
to the parity doubling problem was proposed by Le Yaouanc {\it et
al.}~\cite{yaouanc} in the mid 1980s, i.e. long before any
experimental evidence. Solving the corresponding Bethe-Salpeter
equation with a confining Lorentz-vector potential,
it was observed that the splitting between
the parity doublets decreases as one goes to high masses,
large compared to the scale of spontaneous chiral symmetry
breaking. After appearance of the Crystal Barrel data~\cite{bugg} we can say
that the meson spectrum obtained in~\cite{yaouanc} is in a bad agreement with the
experiment quantitatively, but in a good agreement qualitatively:
Fixing a mass scale large enough, the number of states in the $J^{PC}$ channel $J^{--}$
is a sum of $J^{++}$ and $J^{-+}$ states ($J>0$, $J$ is odd; for $J$ even the situation
is converse), i.e. the right duplication of trajectories was predicted.
As one often says now, an effective CSR was observed.
It is worth noting that only one kind of interpolating currents
was exploited for each channel. Later various modifications of
this model were proposed and a better agreement with the
experimental data was achieved. We refer to the review~\cite{glozrev}
and to~\cite{bicudo} for references and relevant discussions.

The fourth possibility consists in inclusion of derivatives into
interaction vertices preserving the locality of interactions,
\be
L_{\text{int}}(x)\sim g_ij_i(x)F(\partial_x^2)j_i(x).
\ee
where the formafactor $F$ is some polynomial function. The function $F$ can be
chosen so that this extension also describes the "radially"
excited states. This property was demonstrated by Andrianov {\it et al.}
in~\cite{avy,avy2,avy3}. The corresponding models are usually called
quasilocal quark models.
The issue of parity doubling and CSR was later
elaborated in~\cite{aa,wemod,wemod2,wemod3,wemod4}. Within these models,
a clear signal of CSR can be easily demonstrated
analytically for the first meson radial excitations.

Another lines of research have been undertaken for baryons.
L\"{o}ring {\it et al.}~\cite{ml,ml2,ml3,ml4,ml5} developed a relativistic
quark model for baryons on the basis of the three-particle
Bethe-Salpeter equation. Parity doubling within
this picture naturally arose as an instanton-induced effect.

Finally we mention an algebraic rather than effective model for
baryons. Following some ideas from the spectroscopy of diatomic
molecules, Kirchbach {\it et al.}~\cite{kms} constructed a
group-theoretical "rovibron" quark-diquark model describing the Rarite-Schwinger
type of baryon clusters, which we have concerned in Section~4.1.

\subsection{Parity doubling from QCD sum rules}

We have already discussed the underlying idea of emergence of
parity doubling within asymptotic sum rules in Section~4.2.
On the one hand, the OPE dictates a rapid convergence of difference of two-point
correlators for chiral partners at large space-like momenta. On the other hand,
in the large-$N_c$ limit, the meson excitations become narrow and
asymptotic chiral symmetry imposes certain relations among masses
and residues of chiral partners, they are often called sum rules.
In 2000s this issue attracted much attention (see,
e.g.,~\cite{sh,bean,we,weB,weC,cluster,we2,we2b,garda,afon,arriola,golt3}).
An explanation of parity doubling due to a strong suppression of
direct instanton contributions to the two-point correlators
at large space-like momenta was proposed in~\cite{koch}.
In this case the effect is interpreted as a partial restoration
of $U(1)_A$ symmetry.

Beane deduced in~\cite{bean} that the joint constraints of
quark-meson duality in the large-$N_c$ approximation and chiral
symmetry imply degeneracy of excited vector and axial-vector
mesons. According to~\cite{bean}, asymptotically one has for
the masses of parity partners of linearly rising spectrum
\be
m_+^2(n)-m_-^2(n)\xrightarrow[n\rightarrow\infty]{}0.
\ee
A natural question appears, what is the rate of asymptotic CSR?
Later it became clear that the OPE by itself is hardly able to answer this
question. It is an asymptotic expansion at large Euclidean
momenta (it has zero radius of convergence),
which does not contain enough information to provide
an answer. One needs to invoke some additional assumptions besides
the behavior of spectrum at large $n$. Different assumptions
resulted in different patterns of asymptotic rate for CSR.
In~\cite{we,weB,weC,cluster} the exponential decreasing was inferred,
\be
m_+^2(n)-m_-^2(n)\sim e^{-bn},\qquad b>0.
\ee
The polynomial decreasing was obtained in~\cite{sh},
\be
m_+^2(n)-m_-^2(n)\sim \frac{1}{n}.
\ee
The constant behavior is also
possible~\cite{garda,afon,arriola,golt3},
\be
m_+^2(n)-m_-^2(n)\sim \text{const}.
\ee
It must be emphasized that even in the latter case one would see
parity doubling as long as masses are growing, because
\be
m_+(n)-m_-(n)\sim\frac{\text{const}}{m_+(n)+m_-(n)}\xrightarrow[n\rightarrow\infty]{}0.
\ee
We will discuss this case in Section~6.

\subsection{Parity doubling in heavy-light mesons}

Parity doubling in heavy-light quark systems was expected from
simple considerations. Following Bardeen and Hill~\cite{bardeen},
let us perform a gedanken experiment: What happens to the heavy-light
meson spectrum if we could somehow restore the chiral symmetry,
maintaining the other features of confining QCD? One can naturally
expect that the heavy meson masses must remain unaffected in the
first approximation as they arise primarily from the mass of the
heavy constituent quark and the chiral mass gap is rather a
perturbation. In this respect they are qualitatively different
from the ground nucleons. This gedanken experiment suggests that
the explicit chiral $SU(2)_L\times SU(2)_R$ symmetry should somehow
be realized in the heavy meson mass spectrum as long as the heavy
constituent acts as a spectator in the chiral dynamics.
Consequently, even the ground states should appear as approximate
parity doublets. In~\cite{bardeen} this situation was described
by means of a generalization of the NJL model, where the chiral
symmetric and chiral broken phases can be fine-tuned by an
appropriate choice of coupling constant. As a result, the
nearly degenerate chiral partners for the known $(0^-,1^-)$
heavy-light $D$-mesons (the corresponding $(0^+,1^+)$ multiplet)
were theoretically predicted.

A bit earlier Nowak {\it et al.}~\cite{nowak} described the same
phenomenon within a version of constituent chiral quark model of Manohar and
Georgi~\cite{manohar}. The latter can be viewed as a bosonizied
version (i.e. when one formally introduces collective boson
variables constructed from fermion ones)
of the NJL model, which was shown in~\cite{bardeen} and in
some other papers. The mass difference between parity doublets has
a particularly simple form in the given model,
\be
m_{\pm}\simeq m_h\pm m_{\text{con}}, \qquad  m_{\text{con}}\ll m_h,
\ee
where $m_h$ is the bare heavy quark mass and $m_{\text{con}}$ is
the constituent (dynamical) quark mass. Thus,
$m_+-m_-\simeq2m_{\text{con}}$ gives a simple pattern of mass
splitting inside a chiral multiplet due to the chiral mass
gap\footnote{In a formal language we would interpret this result
as follows. Let $|L\rangle$ and $|R\rangle$ be degenerate
left-handed and right-handed eigenstates of parity-invariant
Hamiltonian operator $\hat{H}$. They are related by parity
operator, $\hat{P}|L,R\rangle=|R,L\rangle$, which commutes with
$\hat{H}$. Construct parity-even and parity-odd eigenstates of
$\hat{H}$, $|\pm\rangle=1/\sqrt{2}(|L\rangle\pm|R\rangle)$, which
diagonalize the parity operator,
$\hat{P}|\pm\rangle=\pm|\pm\rangle$. Parity invariance, however,
does not imply equality of masses. A relevant example is the
parity-invariant Hamiltonian (with $\hat{P}$ Hermitian)
$\hat{\tilde{H}}=\hat{H}+\varepsilon\hat{P}$, where the second term is
a perturbation to the Hamiltonian above~\cite{faddeev}. This
perturbation removes the degeneracy, $m_+-m_-=2\varepsilon$.
Parity can be replaced by chirality, and the perturbation
$\varepsilon$ mimics the term with $m_{\text{con}}$ in the Lagrangian
of~\cite{nowak}
differing by sign for chiral partners. Note in passing that the
same example can serve for a formal demonstration of effective CSR.
Let $\varepsilon$ be not small in comparison with energies
of ground $|\pm\rangle$ states. Consider the $n$-th excitations,
$\hat{\tilde{H}}|\pm\rangle^{(n)}=E_{\pm}^{(n)}|\pm\rangle^{(n)}$.
If $E^{(n)}_{\pm}\gg\varepsilon$ the same logic works.}.

In ten years the predicted $(0^+,1^+)$ multiplet of heavy-light mesons
was discovered experimentally. The two groups above wrote the
reminder papers~\cite{bardeen2,nowak2}. The experimental mass
splitting turned out to be even less, $m_+-m_-\simeq m_{\text{con}}$
(for a generally accepted value $m_{\text{con}}\approx300\pm50$
MeV), but the qualitative agreement is undoubtful.

Parity doubling among heavy-light mesons was also studied in~\cite{kalash}
within a version of extended nonlocal NJL model with a confining
instantaneous potential. The effective CSR was demonstrated
analytically for the spectrum of orbitally excited states.
In fact, a similar result was obtained in framework of a more early version
of extended nonlocal NJL model by Bicudo {\it et al.}~\cite{bicudo2}. It
was observed that the spectrum of heavy-light quarkonia becomes
almost parity independent for high spin excitations. This result,
however, was not attached a particular significance.

\subsection{Parity doubling among glueballs?}

In the recent literature a couple of proposals has appeared that
some observed parity doubled states are actually glueballs...

Faddeev {\it et al.}~\cite{faddeev} put forward a geometrical mechanism
for parity doubling of glueballs. It is widely accepted that
glueballs are likely related to closed QCD strings, i.e. to closed
toroidal fluxtubes of chromoelectric field.
These objects can be emitted by a usual long
linear string. Energy of string is proportional to its length,
hence, a closed string should be unstable against shrinkage away
by minimizing its length. On the other hand, within the purely
gluonic part of QCD, the mass gap and color confinement should
prevent such a shrinkage. This implies that there must be
additional contributions to the energy of closed gluonic
string, which is absent for an open string. It is natural to
assume that the source of this stabilizing force is in the
three-dimensional geometry of toroidal configuration.
This configuration is prepared when one bends a finite length open
string and joins its ends. But before joining the ends, the string
can be twisted once around its core, the resulting topology
may prevent from shrinking. The twist can be either
a left-handed or a right-handed rotation around the core.
Thus, degenerate left and right twisted configurations appear
which are related by parity. Experimentally such states could be
prepared from left-handed and right-handed polarized gluons. It
was argued also that in pure Yang-Mills theory these states
could emerge as solitons. The approximate mass degenerate
$\eta_L(1410)$ and $f_0(1500)$ states were advocated to be the first
$(0^{-+},0^{++})$ glueball parity pair.

Kochelev and Min~\cite{koch} applied the instanton mechanism of
partial $U(1)_A$ symmetry restoration to the problem of low mass
glueballs. Analysing the direct instanton contribution to the
difference of two correlators of glueball currents with opposite
parities, they proposed that the recently observed $X(1835)$
resonance is the lowest mass pseudoscalar glueball, which is
parity doubled with the presumably lowest mass scalar glueball
$f_0(1710)$.

Finally, we would mention that another kind of gluonic excitations
has attracted attention recently, the gluelumps. At present, they
do not reveal the parity doubling, nevertheless, these idealized
gluonic constructions are not free from puzzles with the parity as well,
namely they show an unusual ordering of the spin-parity energy
levels (see, {e.g.},~\cite{guo}).

\subsection{Parity doubling and AdS/QCD}

Nowadays a new fashionable approach to quantum field theory has
flourished, the so-called AdS/CFT correspondence (Anti de
Sitter/Conformal Field Theory), which establishes a duality
between string theories defined on the 5-dimensional AdS
space-time and conformal field theories in physical space-time.
The fact that QCD becomes nearly conformal field theory in the
regime, where its effective coupling is approximately constant and
the quark masses can be neglected, i.e., at high momentum
transfer, inspired to apply the AdS/CFT correspondence to QCD,
assuming that QCD can be approximated as a conformal theory
even at relatively small momentum transfer, this conjecture is often
referred to as AdS/QCD approach. Loosely speaking, one tries to
derive the hadron spectrum and strong dynamics from a holographic
dual string theory
defined on five-dimensional AdS space, whose metrics is
a function to be guessed. Such a "bottom-up" approach is often
regarded as a useful (not yet proven) semi-classical approximation
to QCD, which incorporates both color confinement and conformal
short-distance behaviour, see a recent review~\cite{brodsk2} for references.

The AdS/QCD models are believed to provide insights into
non-per\-tur\-ba\-ti\-ve aspects of strongly coupled QCD such as hadron
spectra.
It is natural thus to wonder if this approach may be useful for
the subject in question. To the best of our knowledge, the only
paper which directly addressed the parity doubling within AdS/QCD
is~\cite{brodsk}. The proposed model gives a certain pattern
for the parity doubling among the unflavored baryons with different
angular momenta, which implies a larger symmetry than the
effective chiral symmetry restoration. In a sense, the holographic
models put forward in~\cite{katz,katzf,katzf2} predict the parity
doubling of hadrons as a particular case of the clustering of
resonances expressed by relation~\eqref{hadsp}, the subject of
clustering will be considered in Section~7.

A problem of existing holographic models of QCD is that typically
they lead to the spectrum (see, e.g.,~\cite{sh} for references)
\be
\label{adssp}
m_n^2\sim n^2,
\ee
rather than to the linear one, $m_n^2\sim n$ (the papers~\cite{katz,katzf,katzf2}
are among the exceptions).

We would make a funny observation that exactly the pattern~\eqref{adssp}
was obtained in 1960s (as a particular case) by Barut {\it et
al.}~\cite{bck2,barut3,barut3b} within the dynamical group approach. This
could be regarded as a mere coincidence if it were not a curious
fact that both approaches are essentially based on the conformal
group $O(4,2)$. In Barut {\it et al's.} analysis, this group is
underlying dynamical symmetry, while in holographic duals, the
conformal symmetry is fundamental --- it is taken as a first
approximation to the real-life QCD (where it is broken down by the
conformal anomaly) as long as the gravity/gauge correspondence was
originally established for conformal field theories. Is this a
fortuitous coincidence?

Independently of answer, one might observe some symptomatic
similarities between the impetuous group-theoretical activity of
1960s, which was eager to find a "genuine" spectrum-generating group
for the hadron world,
and the present AdS/QCD (more generally, AdS/CFT) one. In
both cases one looks for a theoretical control over the strong
interactions with the help of some "other" theory, and tries to
find the fittest one (the search for underlying dynamical symmetry
group and its physical representation in one case and the search
for metrics of underlying AdS space and physical boundary conditions
in its holographic variable
in the other). In both cases the activity was inspired by a
successful model example demonstrating a complete realization of
proposed ideology (the hydrogen atom in one case and Maldacena's
example~\cite{maldacena} in the other). In both cases in order to
take into account new real-life features one usually needs more
and more contrived descriptions. The dynamical group approach
finally did not justify ambitious hopes, it
bootstraped itself into complexity (following the fate of
bootstrap models in 1960s) and a large interest to this approach
faded away...

We hope that such historical parallels are premature. Revelation
of parity doubling mechanism is certainly a challenge for the AdS/QCD
models.

\section{Forms of parity doubling}

Up to now, discussing the parity doubling we have skipped a delicate
question: In what situation do we deal with a real parity
doubling, or put it differently, at what mass splitting between states
with opposite parity but equal spin may we say that they are
parity partners? In this respect
the situation with parity doubling in hadron spectrum needs a
further specification as long as in the literature there is no
unique criterium.
%A certain order is needed.
The resonances have a width
which usually grows for highly excited states and eventually the
excitations become practically indistinguishable from the
perturbative continuum. But, at least for mesons, in the limit of
infinite number of colors~\cite{hoof,hoofw}, where the meson states are
infinitely narrow, the problem looks as well defined. We consider
the radially excited states, the case of orbitally excited ones is
similar.

Denote the masses of the $n$-th radial excitations of parity
partners
as $m_+(n)$ and $m_-(n)$. Parity
doubling, in fact, can be understood in different ways. We would
propose the following classification of forms for parity
doubling in the hadron spectrum,
\begin{equation}
\label{clas}
\begin{cases}
m_+^2(n)-m_-^2(n)\xrightarrow[n\rightarrow\infty]{}0 &\text{superstrong;}\\
m_+^2(n)-m_-^2(n)\xrightarrow[n\rightarrow\infty]{}const &\text{strong;}\\
m_+(n)-m_-(n)\xrightarrow[n\rightarrow\infty]{}0 &\text{moderate;}\\
m_+(n)-m_-(n)\xrightarrow[n\rightarrow\infty]{}const &\text{weak;}\\
\dfrac{m_+(n)-m_-(n)}{m_+(n)+m_-(n)}\xrightarrow[n\rightarrow\infty]{}0 &\text{superweak.}
\end{cases}
\end{equation}
Each subsequent definition is less strong than the previous one.
For instance,
consider the mass spectrum where the main asymptotic in $n$ is
linear like in Eq.~\eqref{lin}. The difference of masses squared
can behave as:
\begin{equation}
m_+^2(n)-m_-^2(n)\xrightarrow[n\rightarrow\infty]{}const\times n^{1-\epsilon},\qquad \epsilon>0.
\end{equation}
Then one has according to classification~\eqref{clas}:
$\epsilon>1$ --- superstrong form of parity doubling;
$\epsilon=1$ --- strong
form; $1/2<\epsilon<1$ --- moderate form; $\epsilon=1/2$ --- weak
form; $0<\epsilon<1/2$ --- superweak form. In
papers~\cite{sh,bean,we,cluster} the authors arrived at the {\it
superstrong} form of parity doubling (interpreted as CSR at high
energies). The difference of results in these approaches was in
the estimation of the {\it rate} of CSR. The criterion chosen
in~\cite{golt3} coincides with the {\it superstrong} requirement
if the spectrum for large $n$ is linear.
The spectrum of Ademollo-Veneziano-Weinberg dual amplitude~\cite{gLS}
(a generalized Lovelace-Shapiro amplitude~\cite{LS,LS2}) reveals
the {\it strong} form of parity doubling: The chiral partner
trajectories have an equal slope but different intercepts.
Similar results were obtained in~\cite{kl2,garda,afon,arriola,golt3,katz}
(see also a toy model~\cite{G5} in the chiral broken phase).
Another example of {\it
strong} form is given by the two-dimensional QCD in
a specific sequence of $N_c\rightarrow\infty$ limits, $m_q\rightarrow0$
while $m_q\gg g\sim1/\sqrt{N_c}$
($m_q$ denotes current quark mass and $g$ is coupling constant),
the so-called 't Hooft model~\cite{dim2,dim2b,dim2c} where
the linear spectrum $m^2_n\sim n$ alternates in parity as one increases $n$ by
one unit
(this model, however, has a little to do with the dim4
QCD, see, e.g.,~\cite{glozrev,dim2a,dim2a2}). The same situation takes
place in the models where $n$ is replaced by the "principal
quantum number" of the kind $n=l+2n_r$
(see, e.g.,~\cite{goebel,simonov,simonovb,yamada,kasi,efga,kanesm}) as long as the meson
parity is $(-1)^{l+1}$.
The choice of {\it moderate} form for
parity doubling seems to be natural for the potential and other
nonrelativistic models.
Throughout the review on CSR~\cite{glozrev} the
criterion for effective CSR was adopted in the {\it superweak}
form. A similar assumption was used also in~\cite{pi}.
Various nonlocal extensions of NJL model (see references in~\cite{glozrev,bicudo})
typically reveal the superweak form of parity doubling at low spins
which gradually converts into the superstrong one at high spins.
If the slopes of chiral partner
trajectories turn out to be different (as in the models~\cite{golt,goltc,goltm}
where $\epsilon=0$) one has no parity doubling in any sense.

Different ways of understanding of parity doubling can sometimes lead
to confusing situations. For instance, the 't Hooft model was
treated as an example of parity doubling in~\cite{G4} and as counterexample
in~\cite{golt3,golt2}. The reason is that the superweak form was meant
in the former case and the superstrong  one in the latter.

For the case of the weak and superweak forms the difference $m_+(n) - m_-(n)$
does not converge at all, just it becomes negligible in comparison
with values of masses. This kind of parity doubling can be called
"effective". This is opposed to the "genuine" one which should be defined.
In relativistic theories the latter could be the superstrong form
of parity doubling for bosons. Indeed, if, say, a "genuine" restoration of chiral
and axial symmetries occurs in a part of spectrum, the
corresponding states forget completely about violations of these
symmetries. But parity doubling in the strong form means
that all states are equally influenced by chiral and axial symmetry
breakings
at low energies. In this case we observe parity doubling high in
spectrum because chirally non-invariant contributions to
masses remain constant while the masses are growing,
i.e. low-energy effects equally persist at any energies but become
relatively
unimportant high in energy. In this sense the strong form is also
an "effective" parity doubling. This automatically excludes the
moderate form as a candidate for "genuine" parity doubling in
relativistic theories.

Thus, if reply to the question
"There is or there is no parity doubling in the hadron spectrum?"
is positive, the next question is "What form of parity
doubling is realized in nature?" Are we dealing with a
"genuine" parity doubling or with simply "effective" one?

\begin{center}
\begin{figure}
\vspace{-8cm}
\hspace{-4cm}
\includegraphics[scale=1]{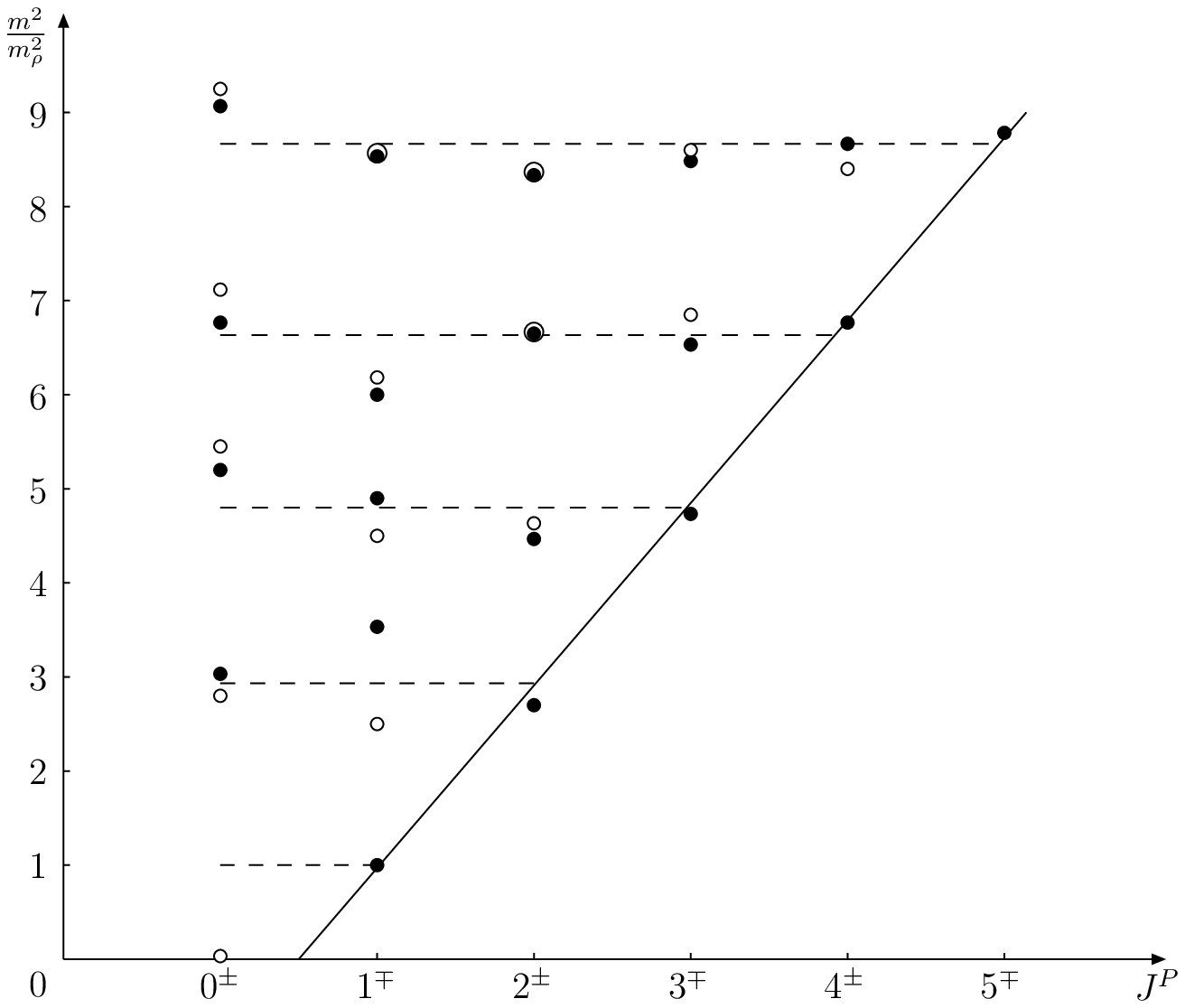}
\vspace{-14cm}
\caption{\small The averaged (masses)$^2$ of states on master
trajectory and on daughter trajectories (filled circles) in units
of $\rho(770)$-meson (mass)$^2$ (see text). The chiral partners are
denoted by open circles. The solid line is the master
trajectory, the dashed lines mark the clustering positions.
The following experimental $J^{P}$ states have been used~\cite{pdg,bugg} (the states discovered in
the Crystal Barrel experiment are marked by star; if a duplication of states happens
(like for some $\rho$- and $\rho_3$-mesons in Fig.~6) the most degenerate chiral partner is chosen).
$0^+$($f_0$-mesons): $1350\pm150$, $1770\pm12$, $2020\pm38(*)$, $2337\pm14(*)$;
$0^-$($\pi$-mesons): $140$, $1300\pm100$, $1812\pm14$, $2070\pm35(*)$, $2360\pm25(*)$;
$1^-$($\rho$-mesons): $776$, $1459\pm11$, $1720\pm20$, $1900\pm?$, $2265\pm40(*)$;
$1^+$($a_1$-mesons): $1230\pm40$, $1647\pm22$, $1930^{+30}_{-70}(*)$, $2270^{+55}_{-40}(*)$;
$2^+$($f_2$-mesons): $1275\pm1$, $1638\pm6$, $2001\pm10(*)$, $2240\pm15(*)$;
$2^-$($\pi_2$-mesons): $1672\pm3$, $2005\pm15(*)$, $2245\pm60(*)$;
$3^-$($\rho_3$-mesons): $1689\pm2$, $1982\pm14(*)$, $2260\pm20(*)$;
$3^+$($a_3$-mesons): $2031\pm12(*)$, $2275\pm35(*)$;
$4^+$($f_4$-mesons): $2018\pm6$, $2283\pm17$;
$4^-$($\pi_4$-mesons): $2250\pm15(*)$;
$5^-$($\rho_5$-mesons): $2300\pm45$.}
\end{figure}
\end{center}

\vspace{-1.1cm}

The Crystal Barrel data~\cite{bugg} allows to offer
a preliminary answer for the
unflavored mesons. Consider the principal $\rho$-meson Regge
trajectory. It consists of $IJ^{PC}$ states $1J^{--}$,
$J=1,3,5\dots$. Consider the principal $f_2$-meson Regge
trajectory. It consists of $IJ^{PC}$ states $0J^{++}$,
$J=2,4,6\dots$. Both trajectories are known to coalesce into one
master trajectory (isospin + exchange degeneracy, see Section~2.3).
Consider the daughter trajectories and the
corresponding chiral partners (the $a_J$ and $\pi_J$ states
respectively). The introduced states are better known experimentally
in comparison with their isospin counterparts, so we confine ourselves
by the states above. The known averaged masses squared are
depicted in Fig.~7 in units of $\rho(770)$-meson mass squared.
Master trajectory
is known to be populated by the parity singlet states only. As to
the differences of (masses)$^2$ between chiral partners, there is
a certain tendency to converge systematically high in the spectrum.
This convergence means that nature seems to prefer the superstrong
form of parity doubling among light mesons.
In addition, the resonances apparently cluster near approximately
equidistant values of masses squared (approximately near $1.3^2$,
$1.7^2$, $2^2$, and $2.3^2$ GeV$^2$), this issue will be discussed
in the next Section.

Let us believe that the superstrong form is indeed realized. The
next qualifying question is then "What is the actual {\it rate}
of parity doubling?" We have already concerned this issue in
Section~5.3 from the theoretical side. But what about experiment?
As a rough estimate, we can propose an averaged
splitting between (masses)$^2$ of chiral partners within each
cluster. Define the averaged splitting as
\be
\label{delt}
\Delta_i=\frac{1}{N_i}\sum_{k=1}^{N_i}\frac{|m_{ik,+}^2-m_{ik,-}^2|}{m_{\rho(770)}^2},
\ee
where $N_i$ is the number of chiral pairs inside the $i$-th
cluster (in fact, they are related by $N_i=i$ for $i>1$, see Fig.~7), $i$ enumerates
the clusters in increasing energy, $i=1,2,3,4,5$
($i=1$ corresponds to the lowest cluster where there is no
chiral pair), and $m_{ik,\pm}$ stays for the mass of $\pm$ state in the $k$-th
chiral pair of the $i$-th cluster. The mass of $\rho(770)$-meson
can be regarded as QCD mass gap. Thus, the quantities
$\Delta_i$ "measure" averaged chiral symmetry breaking effects at
different energy scales
in the units of QCD mass gap. We will regard also the lowest $\pi$ and
$\rho$ mesons as "would be nonlinear" chiral pair. The deviation
of quantity $\Delta_1$ from unity "measures" then an explicit chiral
symmetry breaking in the QCD Lagrangian (of the order of
$m_{\text{current}}/m_{\text{constituent}}$).

The results of our "measurements" for the states in Fig.~7 are
\be
\Delta_1\approx0.97,\quad
\Delta_2\approx0.62,\quad
\Delta_3\approx0.28,\quad
\Delta_4\approx0.22,\quad
\Delta_5\approx0.12.
\ee
The series of numbers \{0.97, 0.62, 0.28, 0.22, 0.12\} seems to be
convergent. Keeping in mind the discussions in Section~5.3,
one could ask a question of the kind "What continuous function does
interpolate this series in the best possible way?" We think that
any answer to such a question is hardly able to help to the parity
doubling problem from the theoretical point of view.

A more constructive question may be the following. If the
superstrong form is indeed realized, the mass splitting within
chiral pairs can rapidly become less than the typical experimental
errors in determination of masses. Suppose that parity doubling
occurs due to internal symmetries (chiral and axial).
The experimental situation is then
indistinguishable from the explicit Wigner-Weyl realization of
chiral symmetry (provided that pions are decoupled). This means a
complete restoration of chiral (and axial) symmetries of the
classical QCD Lagrangian in the
hadron spectrum. Thus, what is a scale of this restoration?
We are aware of three such estimations in the literature:
from a kind of nonlocal extension of the NJL model~\cite{kalash},
from a phenomenological analysis~\cite{afon2} (the position of
"would be" the next cluster in Fig.~7), and from analysis of
decreasing of the constituent quark mass in response to increasing
of momentum~\cite{swan}. These three very different approaches
remarkably converged to one number: 2.5~GeV. Unfortunately, we do
not know whether the unflavored meson resonances persist systematically
at such an energy scale. It may be that the perturbative QCD
continuum sets in, and CSR becomes trivial. The Crystal Barrel
experiment measured the relevant resonances up to 2.4~GeV only.
{\it One certainly needs new experiments devoted to the search for new
meson resonances, which cover the range at least up to 2.6~GeV.
These extra 200 MeV would be decisive in checking various proposals
about the global features of meson spectrum.}

Let us summarize the distinctive global features of unflavored
meson spectrum which attracted attention recently:\\
1) The systematic appearance of parity doublets (presumably everywhere
except pion and the states belonging to the master trajectory).\\
2) The systematic appearance of parity singlets --- they occupy
completely the master trajectory.\\
3) The states with different spins cluster near certain values of
energy.

We propose to compare Fig.~7 with Fig.~1 (a hydrogen-like assignment
of hadron levels within the dynamical $O(4,2)$ conformal group).
The same pattern of spectral degeneracies was suggested for baryons
in the preQCD time!

Let us pass on to baryons. The mass of ground nucleon is known
to be mostly induced by the chiral symmetry breaking~\cite{ioffe,ioffeb}.
Consider the quantities $\Delta_i$ defined as in Eq.~\eqref{delt}
with the replacement
\be
m_{\rho(770)}\rightarrow m_{N(939)}.
\ee
Thus, we will "measure" parity doubling in the units of {\it
chiral} mass gap in QCD. The first cluster will be simply the
ground nucleon state.
As fermion masses enter relativistic
equations linearly, it may be useful to consider also an analogous
to $\Delta_i$ quantities for linear masses,
\be
\label{delt2}
\delta_i=\frac{1}{N_i}\sum_{k=1}^{N_i}\frac{|m_{ik,+}-m_{ik,-}|}{m_{N(939)}}.
\ee
A convergence of $\delta_i$ signals that at least the weak
form of parity doubling is realized (or the moderate one for the convergence to
zero). We will also check the superweak form of parity doubling
in the non-strange baryons with the help of the quantities
\be
\label{delt3}
\chi_i=\frac{1}{N_i}\sum_{k=1}^{N_i}\frac{|m_{ik,+}-m_{ik,-}|}{m_{ik,+}+m_{ik,-}}.
\ee
The mass values for parity partners are contained in the Particle
Data~\cite{pdg} (see also~\cite{glozrev}). The results of our
"measurements" are summarized in Table~1.

\begin{center}
\begin{table}
\caption{\small The values of quantities $\Delta_i$, $\delta_i$, and $\chi_i$ (see text)
for the experimental spectrum of nucleons and deltas~\cite{pdg}.}
\smallskip
\begin{center}
\begin{tabular}{c@{\hspace{0.8cm}}cc@{\hspace{0.8cm}}cc@{\hspace{0.8cm}}cc}
\hline
\hline
&\multicolumn{2}{c}{$\Delta_i\phantom{\Delta\Delta}$}&\multicolumn{2}{c}
{$\delta_i\phantom{\Delta\Delta}$}&\multicolumn{2}{c}{$\chi_i\phantom{\Delta}$}\\
i&$N$ & $\Delta$&$N$ & $\Delta$&$N$ & $\Delta$\\
\hline
1&1&&1&&1&\\
2&0.32&0.41&0.10&0.11&0.032&0.032\\
3&0.10&0.15&0.03&0.04&0.008&0.009\\
4&0.58&0.76&0.13&0.15&0.030&0.031\\
\hline
\hline
\end{tabular}
\end{center}
\end{table}
\end{center}

\vspace{-0.6cm}

It can be readily seen that without account for the highest
cluster one likely has the superstrong form of parity doubling.
However, the last cluster spoils this nice picture --- there is a
vague hope for the superweak form, at best. Qualitatively the
deterioration of parity doubling in the highest known non-strange
baryons can be foreseen from a glance at Fig.~3 and Fig.~4. On the
other hand, these figures show visually how poor is the
experimental determination of masses in the highest clusters.
The spectral regularities cease to be clear-cut.
Thus, {\it one is in great need of high precision experiments
both for determination of masses of unflavored baryons and for
search for new states}.

\section{The latest idea: Broader degeneracy?}

We have already noted several times that the non-strange hadrons
tend to cluster into fairly narrow mass range (see discussions
on meson "towers" in Section~2.3, on H\"{o}hler's clusters in
Section~4.1, and on qualitative features of meson spectrum in relation
to Fig.~7 in Section~6). Clustering of unflavored mesons was
clearly observed in Crystal Barrel experiment (see Fig.~4
in~\cite{bugg}). A preliminary picture for non-strange meson
spectrum, as we know it now, is displayed in Fig.~8.

\begin{center}
\begin{figure}
\vspace{-5cm}
\hspace{-4cm}
\includegraphics[scale=1]{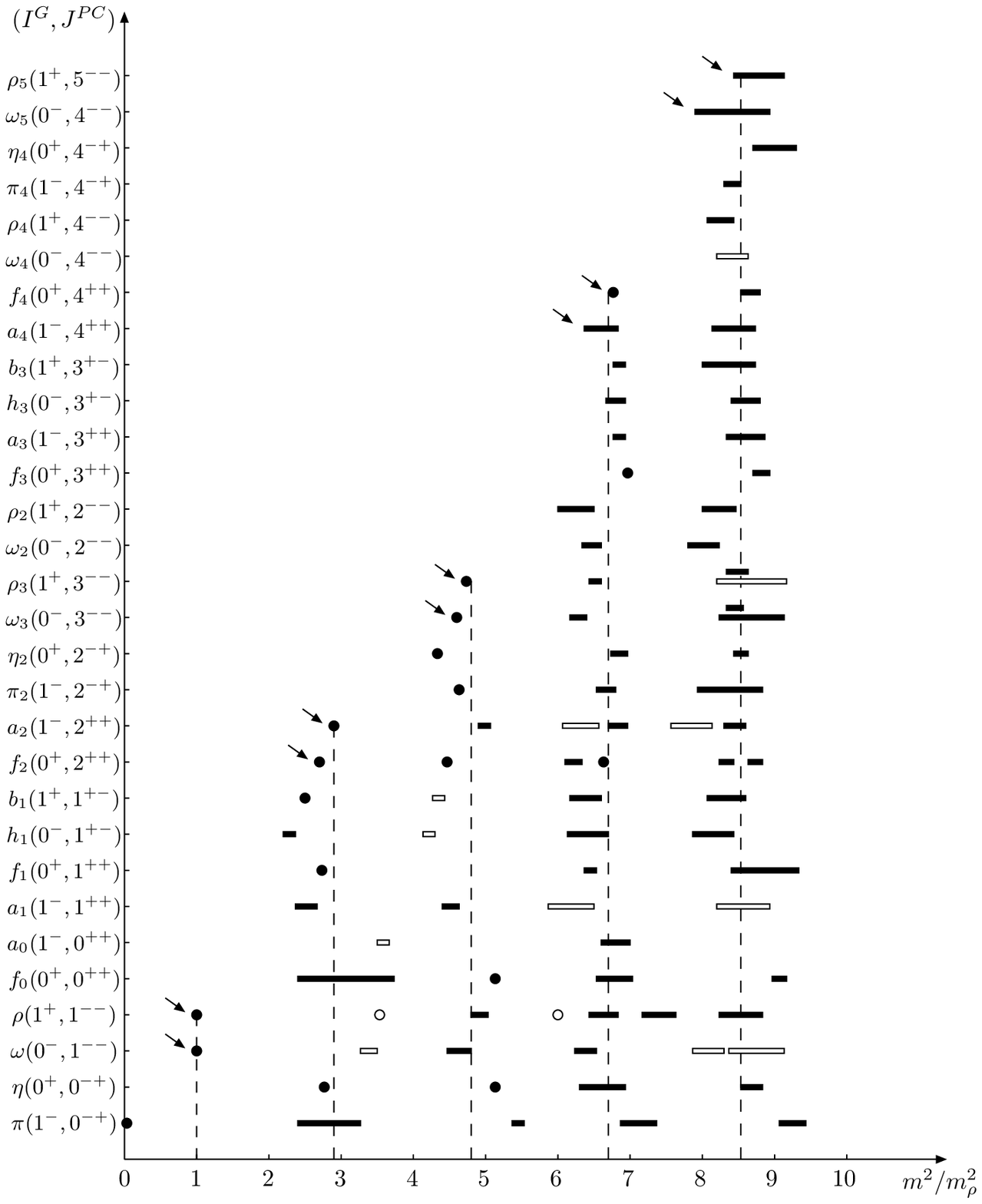}
\vspace{-8cm}
\caption{\small The spectrum of non-strange mesons in units of $\rho(770)$-meson
mass squared (plot from~\cite{afonlatest}).
The data for the first three clusters is contained in Particle Data~\cite{pdg},
while the data for the last two ones is mostly due to Crystal Barrel~\cite{bugg}.
Experimental errors are indicated.
Circles stay when errors are negligible. The dashed lines mark the
mean (mass)$^2$ in each cluster. The open strips and circles
denote either
the states with the lowest star rating according to~\cite{bugg} or the states which
are somewhat doubtful as candidates for the unflavored mesons.
The arrows indicate the $J>0$ mesons which have no chiral partners
(the candidates for chiral singlets).}
\end{figure}
\end{center}

\vspace{-1.2cm}

The clustering is a striking phenomenon which theoretically is
even more enigmatic than parity doubling. Why do many states with
different spins and other quantum numbers have close masses?
In the last year this subject has been attracting much
attention~\cite{glozrev,afonepj,bicudo,afon2,afonlatest,cohen,shvain,prc}.
As a matter of fact, in some approaches parity doubling is a mere
consequence of some kind of clustering (see, e.g., Kirchbach's and Klempt's
models discussed above).
If some force drives hadron masses to cluster into well separated
narrow mass regions then parity doubling ensues automatically
provided that opposite parity states exist systematically in the
resonance region. In this case the mass splitting between parity
(chiral) partners is of the order of "width" of narrow mass region
occupied by given cluster. A surprising observation is that
exactly this picture seems to be the case experimentally.

The present experimental data shows three qualitative distinctions
between the baryon clusters in Fig.~3, Fig.~4 and the meson clusters
in Fig.~8 (or its abridged form, Fig.~7). First, the meson
clusters are approximately equidistant, while this is not the case
for the baryon ones. Second, the baryon clustering becomes much worse
high in energy, while the meson one is progressively improving up
to the highest available resonance region. Third, all meson states
belonging to the lowest principal Regge trajectories are parity
singlets (they are indicated by arrows in Fig.~8), the baryon
sector has several exceptions from this rule.

If the linear parametrization of meson spectrum~\eqref{lin} is
approximately valid with the universal slope $a$, then clustering
in the meson sector is essentially equivalent to an approximate
universality of the intercept $m_0^2$. What is the averaged value
of $m_0^2$ in the units of $a$? This question was addressed
in~\cite{afonlatest}. Averaging over all states in Fig.~8,
the value is $m_0^2/a\approx1/2$, like in the spectrum of
Lovelace-Shapiro dual amplitude~\cite{LS,LS2}. More precisely, the
following averaged spectrum was obtained for the states in Fig.~8
\be
\label{hadsp}
m^2(l,n_r)=a(l+n_r+c).
\ee
with $a\approx1.1$ GeV$^2$ and $c\approx0.6$.
Compare the classical hydrogen spectrum, Eq.~\eqref{hydsp}, with
the approximate meson spectrum, Eq.~\eqref{hadsp} (see also
Klempt's formula, Eq.~\eqref{lin2}). The outcome is
that a broad degeneracy emerges due to the existence of single "principal"
quantum number
\be
\label{princ}
n=l+n_r+\text{const}.
\ee
The validity of such nonrelativistic relations
presupposes a smallness of spin-orbital interactions. Recalling
the nonrelativistic definition of parity for quark-antiquark
pair, $P=(-1)^{l+1}$, relation~\eqref{hadsp} immediately
reproduces the absence of parity doubling for the leading meson
trajectories: The corresponding states have $n_r=0$ and the minimal
value of $l$ at given spin, while the closest parity partners have
$l$ larger by one unit (hence, according to Eq.~\eqref{hadsp},
they are degenerate with the first "radial" excitations of the states
lying on the principal trajectories). This reasoning does not work
for baryons if Eq.~\eqref{lin2} is valid --- the intercept is not
universal, thus, parity doubling on principal trajectories is not
forbidden.

The clustering constitutes, perhaps, a problem of paramount importance
for modern spectroscopy of non-strange hadrons. What is the
underlying physics?
The restoration of chiral and axial symmetries cannot lead to
the multispin clustering since the corresponding transformations
relate states with equal spin only. Many other approaches put
forward for explanation of parity doubling have also problems with
a natural accommodation of clustering. The hadron strings are somewhat
encouraging (see, e.g.,~\cite{baker} where Eq.~\eqref{hadsp} was
qualitatively derived), but a consistent relativistic theory of hadron
strings predicts plenty of unobserved "hybrid" states, not to
mention the unresolved problem with tachyons in four dimensions.
It would be interesting to address the problems of parity doubling
and clustering within the framework of presently fashionable
AdS/QCD approach, for the time being only a few of holographic
models reproduced a spectral pattern in light mesons~\cite{katz,katzf,katzf2},
which is similar to the observed one, Eq.~\eqref{hadsp}.
The description of large degeneracy is a challenge for quark
models, recently this problem was emphasized by
Bicudo~\cite{bicudo}. The most ambitious approach in this field
is to find a quark model with a principal quantum number, at least
in some approximate sense. The existence of effective principal quantum
number in the spectrum is a strong argument in favor of the hydrogen
like classification of
unflavored mesons, which was put forward recently~\cite{afonrec}.
The underlying qualitative motivation for such a classification is
that both the hydrogen atom and the mesons
represent quantum two-body systems interacting via central forces,
so it looks plausible that they could have some general dynamical
symmetries.

The dependence of spectrum on one quantum number only, Eq.~\eqref{princ},
may be regarded as a compact form to express the combined effect
of suppression of spin-orbital and exchange forces. To explain the
statement, consider a $J^-$ state on leading Regge trajectory $R_-$. If
the spin-orbital interaction is small, the state is approximately
degenerate with $(J-1)^-$, $(J-2)^-$, and so on, in the baryon
case we referred to such a tower as Feynman's cluster (see
Section~4.1). The same can be repeated for a $J'^+$
state on leading Regge trajectory $R_+$. However, due to exchange
degeneracy (see Section~2.3), the trajectories $R_-$ and $R_+$
coincide. Consequently, the $(J-1)^-$ state will be degenerate
with the $(J'-1)^+$ lying on the first daughter of $R_+$ (the
first "radial" excitation). This chain can be continued, and as a
result one obtains parity doublets for all daughter trajectories.
The described mechanism produces the observed pattern of parity
doubling among mesons, see Fig.~7.

There is a hope to obtain this mechanism from confinement in
QCD~\cite{kaidalov2}. If the confinement dynamics follows from
the area law for large Wilson loops, then linearity and other
properties of
Regge trajectories can be derived under some assumptions (see,
e.g.,~\cite{kaidalov3,kaidalov3b}). Along this line, Eq.~\eqref{hadsp} was
derived in~\cite{simonov2} in a model-dependent way for large
angular momentum, in this limiting case the interquark separation
is large and, hence, only the large area asymptotics for Wilson
loops seems to be essential. Similar results were obtained also
for baryons~\cite{simonov3} assuming a large quark-diquark
separation. The existence of single quantum number, Eq.~\eqref{princ},
is expected due to hidden conformal invariance resulting from
reparametrization invariance for Wilson loops~\cite{kaidalov2}.

Thus, at present there is no understanding whether the observed large
degeneracy (and parity doubling as a particular case) is related
with fundamental symmetries of QCD or with some dynamical
symmetry. It may be that the separation itself between
"fundamental" and "dynamical" symmetries is somewhat artificial.
For instance, the chiral $U(n_f)\times U(n_f)$ invariance of QCD
Lagrangian (in the chiral limit) is broken down to the diagonal
$U(n_f)$ subgroup, which is a symmetry of spectrum. The
"remainder" is realized dynamically through massless Goldstone
bosons, thus, giving rise to all low energy dynamics. In this
sense the fundamental chiral symmetry is converted into the
dynamical one~\cite{wein2}. Initially, QCD possesses a mass gap,
but the dynamical realization of chiral symmetry removes a ban,
and generates gapless Goldstone excitations. Higher in energy the
fundamental chiral symmetry, likely, gradually restores and not
far from the onset of complete restoration (the perturbative
continuum) the hadron spectrum reveals the fundamental chiral
symmetry through parity doubling. One could imagine the following
rather unorthodox situation. The conformal $O(4,2)$ invariance of
QCD Lagrangian (in the chiral limit) is broken down to the maximal
compact subgroup $O(4)$, which is a symmetry of spectrum. The
"remainder" is somehow realized dynamically. Initially, the
fundamental conformal symmetry prohibits any massive excitations
--- spectrum of conformal theories is massless or continuum ---
but a dynamical realization of conformal symmetry removes a ban,
and massive states are allowed. Higher in energy the fundamental
conformal symmetry, likely, gradually restores and not far from
the onset of complete restoration (the perturbative continuum) the
hadron spectrum reveals the fundamental conformal symmetry through
clustering --- the gradual restoration simply means a gradual
cleaning from "singled out" regions in the energy distribution,
a liberation from enhancements of any kind, the resonances represent
these very regions, and hadrons are thereby "expelled" in
the resonance region forming narrow clusters, like Abrikosov
vortices are formed in a type-II superconductor when the strength
of magnetic field lies between the first and second critical
values. These values would correspond to the scales of breaking
(mass gap) and restoration of conformal symmetry in this analogy
as at the first critical value the magnetic vortices appear inside
a type-II superconductor while at the second one the
superconductivity is destroyed. Unfortunately, all such suggestive
analogies are highly speculative at present stage, so it is worth
finishing them here.

Descending to down-to-earth discussions, it should be mentioned that
the results of the Crystal Barrel experiment have not yet been
confirmed convincingly, this uncertain situation gives rise to
a rather widespread
opinion that all related discussions, like parity doubling and
clustering in mesons, do not have a solid ground --- it is not excluded that the
systematic character of the observed effects is absent at all. Such
a point of view may be correct when discussing some separate
unconfirmed states, but it is hardly correct when discussing
clustering of states. It seems to be timely to remind some forgotten
experiments.

The clusters are observed as peaks in differential cross sections. A
high-precision experiment is able to distinguish the saw-tooth
structure of these peaks, which depends on a concrete reaction.
The determination of separate resonances then follows. The point
is that peaks like in Fig.~8 with close positions were observed many
times in 1960s, but the instrumental resolution usually did not allow to
distinguish separate resonances. Perhaps, the most prominent old
experiment devoted to the search for missing resonances is the one
performed with the use of the CERN missing-mass spectrometer in the
mid 1960s~\cite{focacci}.
The charged non-strange bosons $X^-$ were produced in the reaction
$\pi^-+p\rightarrow p+X^-$. Mass spectra were obtained by
measuring the missing-mass of the recoil protons in the range
0.5 - 2.5 GeV. Apart from the known at that time peaks (in MeV)
$\rho(768\pm5)$ and $a_2(1286\pm8)$, the following peaks were
observed: $R(1691\pm30)$ (consisting of three separate peaks),
$S(1929\pm14)$, $T(2195\pm15)$, and
$U(2382\pm24)$. We have already mentioned the mass regions $R$,
$S$, $T$, and $U$ in Section~2.3 in relation to another
experiments (see references in~\cite{bar2}). A remarkable
observation was that the masses square $M_X^2$ of the $\rho$,
$a_2$, $R$, $S$, $T$, and $U$ regions lie on a straight line!
The slope turned out to be 1.05 GeV$^2$, this value was very close
to the baryon slope 1.04 GeV$^2$ known at that time (such
observations resulted in the idea of dynamical supersymmetry
between baryons and mesons~\cite{anselmino}).
Even more forgotten is the fact that this experiment was continued
with the CERN boson spectrometer exploiting the same
reaction. At the first stage, the mass region 2.5 - 3.0 GeV was
measured. Two peaks were observed, at 2.62 and 2.80
(with a close peak at 2.88) GeV~\cite{baud1}. The
corresponding masses square happened to lie on the extrapolated
$\rho-a_2-R-S-T-U$ linear trajectory! At the second stage,
the mass region 3.0 - 3.8 GeV was investigated. The spectrum
showed four prominent peaks at 3.025 (with a close peak at 3.075),
3.145 (with a close peak at 3.180), 3.475, and 3.535 GeV~\cite{baud2}.
A less significant peak at 3.605 GeV was detected as well. The
compiled spectrum is displayed in Fig.~9. Some of the discovered peaks
were observed also by other experiments. For instance, near 3.01 GeV in
a inelastic $\pi p$ reaction~\cite{yost}, near 3.03 and 3.4 GeV in
a $\bar{p}p$ annihilation~\cite{alexander}, Particle
Data~\cite{pdg} cites (in section "Further States") a resonance
structure of unknown quantum numbers near mass regions 2.38, 2.62, and
3.02 GeV, which was produced in various inelastic $\pi p$ and $\bar{p}p$
reactions.

\begin{center}
\begin{figure}
\vspace{-5cm}
\hspace{-4cm}
\includegraphics[scale=0.9]{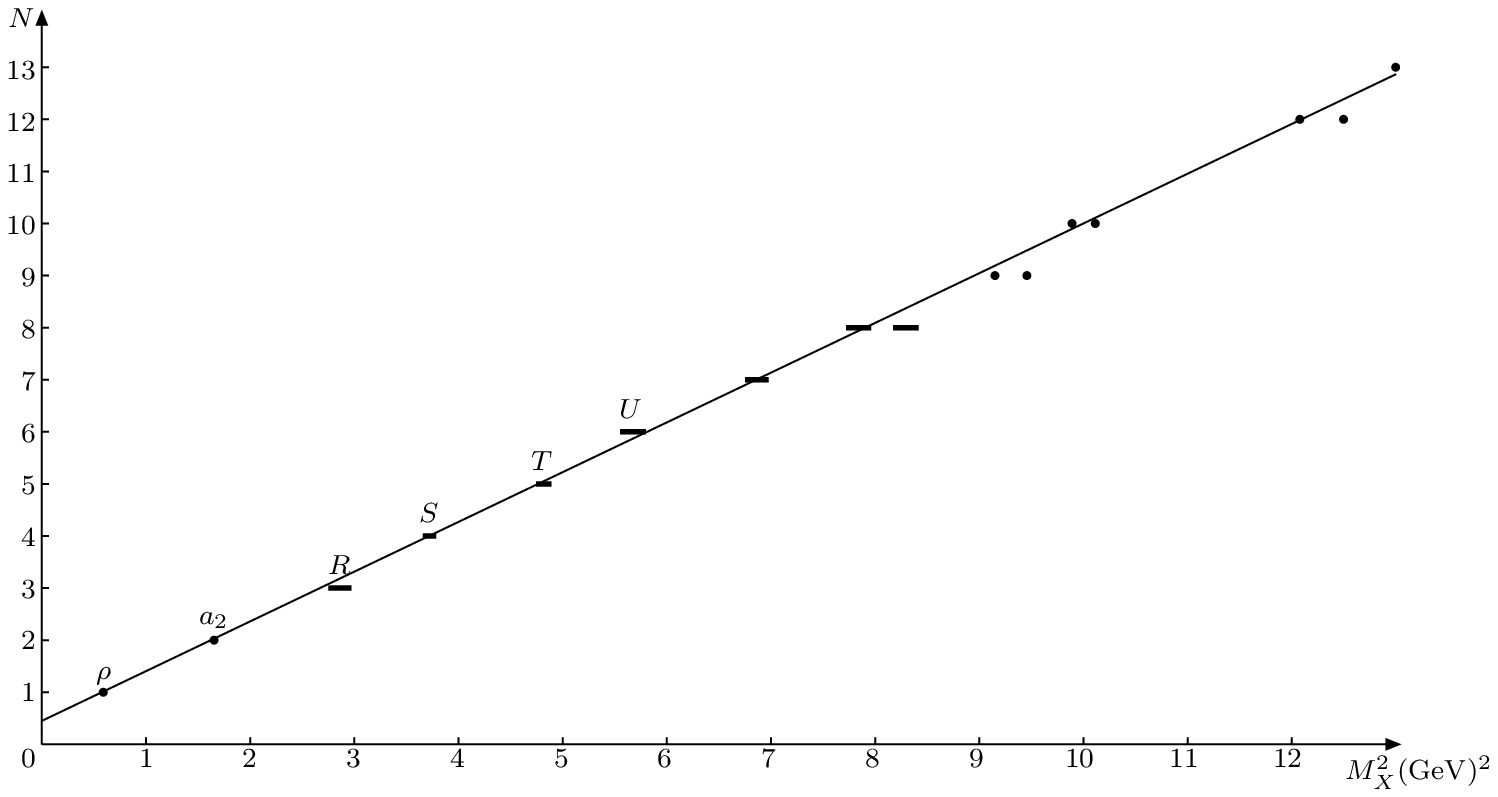}
\vspace{-16.5cm}
\caption{\small The spectrum observed in the CERN missing-mass experiments in 1960s (see text).
$N$ is the peak number.}
\end{figure}
\end{center}

\vspace{-1.3cm}

The plot displayed in Fig.~9 suggests that the resonance region is not
limited by the $U$-region, it extends at least up to 3.6~GeV. In
addition, the linear behavior of objects in Fig.~9 hints at
some Regge-like recurrence, possibly a recurrence of meson
clusters.
It would be highly desirable to recommence such experiments using
the progress in resolution technics achieved for the last 40 years.

\section{Conclusions}

We have tried to trace the development of parity doubling ideas
in particle physics since 1960s and up to the latest publications.
The history of parity doubling is tightly interrelated with the
history of strong interactions. The development of many approaches
invented for description of strong interactions may be looked at
from the point of view of parity doubling. For instance, the
evolution of effective quark models has passed two stages. At the
first stage the quark models were not able to reproduce the
systematic parity doubling. At the second stage, more refined
models naturally accommodated the phenomenon. It seems that soon
we will be witnesses of the third stage --- the creation of models
which describe the clustering of hadrons at certain energies, i.e., a
broader degeneracy than parity doubling.

There is still no agreement whether parity doubling is a
reflection of approximate classical symmetries of QCD or it is a
dynamical effect emerging due to certain internal space structure
of hadrons. Interpreting parity doubling as a manifestation of
some fundamental symmetries, one encounters a bifurcation point
--- are they space-time or internal symmetries? A delicate problem
of two-flavor sector is that these alternatives turn out to be somewhat
dual from the group-theoretical point of view: Extending both
the Lorentz and the chiral group by parity, one obtains a group
isomorphic to $O(4)$, where the irreducible representations
for parity eigenstates are given by
$(j_1,j_2)\oplus(j_2,j_1)$ if $j_1\neq j_2$.
This circumstance hampers to reveal the genuine character of
underlying symmetry.

We have demonstrated that experimentally parity doubling
in non-strange baryons and mesons is not of the same type. May be
this is an artefact of insufficient experimental data. The
deficiency of reliable experimental information is a serious
problem in modern
spectroscopy of non-strange hadrons. It would be desirable if
parity doubling in baryon and meson sector had the same origin. In
this case one obtains a powerful selective principle: Any model
for parity doubling which treats baryons and mesons differently
is missing essential physics, hence, it has to be ruled out.
The idea of approximate meson-baryon dynamical supersymmetry is
suggestive in this respect. Unfortunately, one does not have enough
arguments to postulate that principle.

To summarize, at present the issue of parity doubling has much more
questions than ready answers and definite conclusions. In the
review we have tried to convince in the extreme importance of
experimental searches for missing states in the non-strange
sector and confirmation of preliminarily known states.
In this respect it should be added that presently the lattice
simulations are not able to shed light on the spectrum of high
excitations, hence, on parity doubling. It should be added also
that a more precise specification of light hadron spectrum
requires the rather modest amounts of money and resources.
In particular, a rich experimental
information is accumulated at Jefferson Lab (TJNAF), a careful
analysis of this data could be invaluable for the spectroscopy of
excited nucleons and deltas. The same can be said about the meson
spectrum near 1.7~GeV, which could be finally established using
the VES and E852 data. The future experiment of the PANDA
Collaboration at GSI could refine and extend significantly
the spectroscopic results of the Crystal Barrel one on unflavored
mesons. The same task would be accomplished, at least partly, in a
closer perspective if a new polarised target were used inside the
old Crystal Barrel detector (or any other detector, e.g.,
the Babar detector after its present experiment ends). All this is
quite realizable, a good will is needed only.
We hope that future experiments will give
the long-awaited whole and precise picture of light hadron spectrum, thus
providing a key for ultimate explanation of the parity doubling
phenomenon as well as other spectroscopic puzzles.

\section*{Acknowledgments}
I would like to thank the participants of XLI PNPI Winter School
on Nuclear and Particle Physics for stimulating discussions,
especially A. B. Kaidalov for reading the manuscript and
enlightening comments. The correspondence with D. V. Bugg and
M. Kirchbach is gratefully acknowledged.
The work was supported by RFBR,
grant 05-02-17477, by the Ministry of Education of Russian
Federation, grant RNP.2.1.1.1112, and by grant LSS-5538.2006.2.

\end{document}